\documentclass[reprint, aps, pre, superscriptaddress,showpacs]{revtex4-1}

\usepackage{subfigure}
\usepackage{epsfig}
\usepackage{amsmath}
\newcommand{\D}[2]{\frac{\partial #1}{\partial #2}}
\newcommand{\DD}[2]{\frac{\partial^2 #1}{\partial #2^2}}
\newcommand{\Dstraight}[2]{\frac{\ud #1}{\ud #2}}
\newcommand{\DDstraight}[2]{\frac{\ud^2 #1}{\ud #2^2}}
\newcommand{\ud}{\mbox{d}}
\newcommand{\bs}[1]{\boldsymbol{#1}}
\usepackage{ifpdf}
\ifpdf	
	\newcommand{\ignoreThis}[1]{}
	\usepackage{color}	
	\definecolor{Gray}{rgb}{0.6,0,0}
\else		
	\newcommand{\ignoreThis}[1]{#1}
	\usepackage[usenames,dvips]{color}
\fi

\begin{document}
\title{Onset of collective motion in locusts is captured by a minimal model}
\author{Louise Dyson}
\thanks{These authors contributed equally to this work.}
\affiliation{Mathematics Institute, University of Warwick, Coventry CV4 7AL, UK}
\affiliation{Theoretical Physics Division, School of Physics and Astronomy, University of Manchester, Manchester M13 9PL, United Kingdom}

\author{Christian A. Yates$^*$}\email[Corresponding Author: ]{c.yates@bath.ac.uk}
\affiliation{Centre for Mathematical Biology, Department of Mathematical Sciences, University of Bath, Claverton Down, Bath BA2 7AY, United Kingdom}

\author{Jerome Buhl}
\affiliation{School of Agriculture, Food and Wine, Waite Building, Waite Campus, The University of Adelaide, Adelaide, South Australia 5005, Australia}
\author{Alan J.~McKane}
\affiliation{Theoretical Physics Division, School of Physics and Astronomy, University of Manchester, Manchester M13 9PL, United Kingdom}
\begin{abstract}
We present a minimal model to describe the onset of collective motion seen when a population of locusts are placed in an annular arena. At low densities motion is disordered, while at high densities locusts march in a common direction, which may reverse during the experiment. The data is well-captured by an individual-based model, in which demographic noise leads to the observed density-dependent effects. By fitting the model parameters to equation-free coefficients, we give a quantitative comparison, showing time series, stationary distributions and the mean switching times between states.
\end{abstract} 

\maketitle

\section{Introduction}
Locusts and other migrating insects can form cohesive swarms at large population densities, which subsequently travel over huge distances and can have a devastating effect on agriculture. It is therefore important to understand the mechanisms governing how the population decides collectively on the direction of migration, and the population density at which this occurs. Stochastic models of collective migration can be used to study the population-level effects of individual-level decisions, and can demonstrate sudden changes in collective motion at increased group sizes~\cite{vicsek1995ntp,czirok1999cms,biancalani2014nib,huepe2011adaptive}. Investigating the information that an individual may have within a population is an area of active research and is used in modelling efforts~\cite{pearce2014rpc,bialek2012smn}. Although many existing models produce motion which is qualitatively similar to a variety of forms of collective behaviour~\cite{strombom2011cmf,czirok1999cms,vicsek1995ntp,partridge1982sff,partridge80,Reynolds87,stocker1999mtf, ballerini2008ira,yates2009inc} (swarming, schooling, flocking etc.), very few provide a quantitative comparison to experimental data~\cite{couzin2003soa,sumpter2008itm,berdahl2013esc,mann2013msi,bode2010lif}.
 
\begin{figure}[h!!!!!!!!]
\begin{center} 
\subfigure[N = 5]{
\includegraphics[width=0.29\columnwidth]{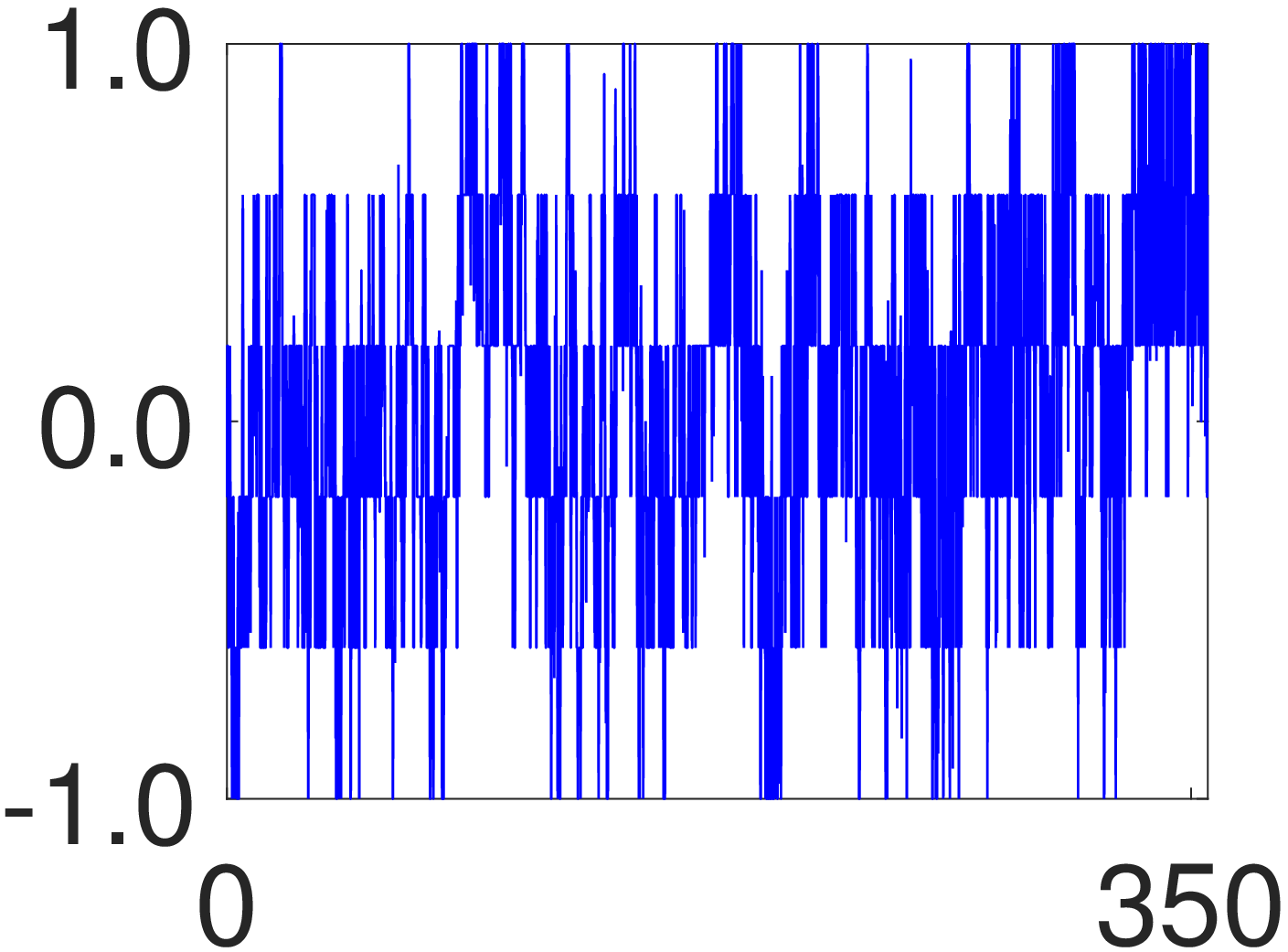}
\label{figure:5_locusts}
}
\subfigure[N = 20]{
\includegraphics[width=0.29\columnwidth]{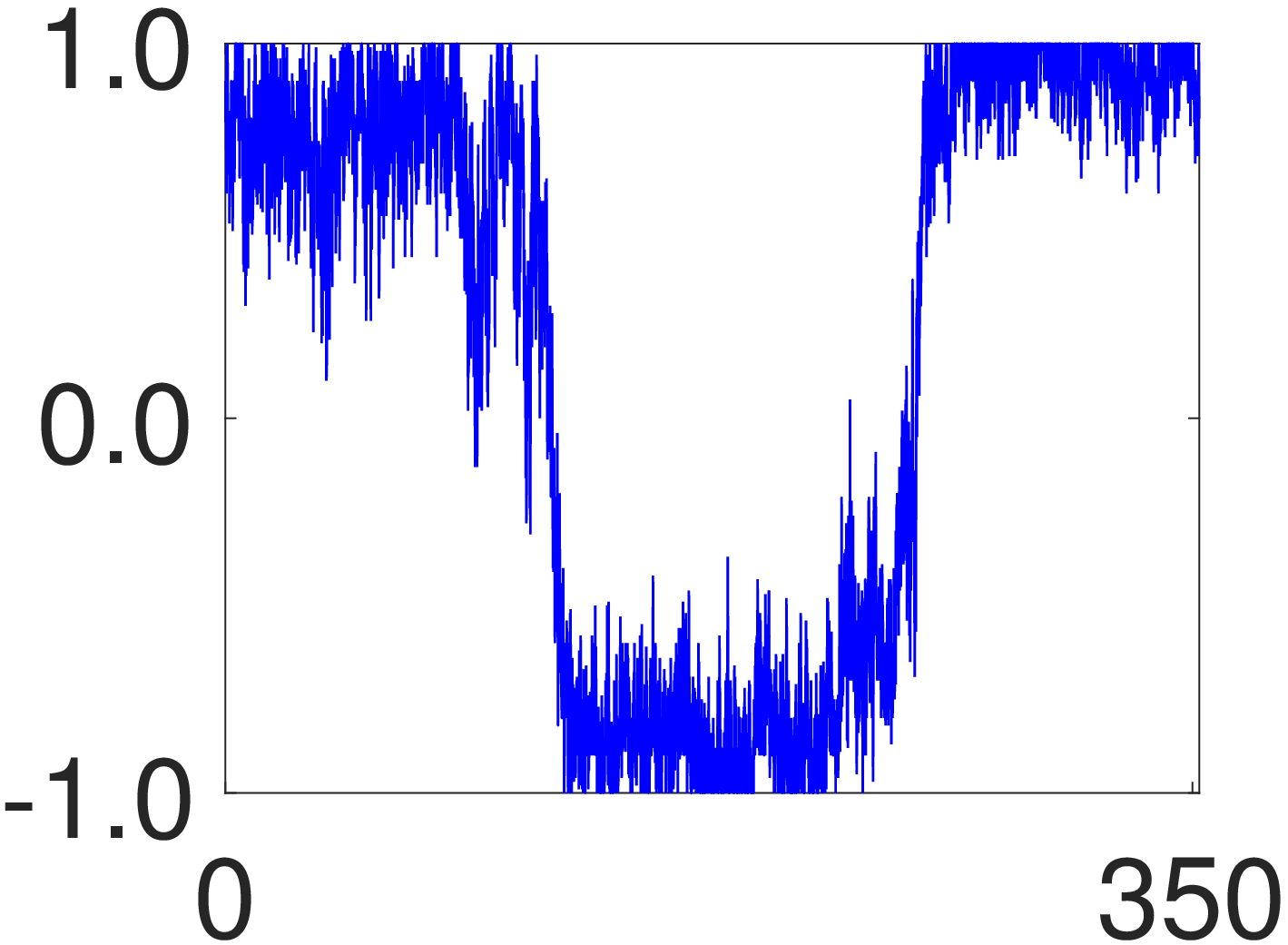}
\label{figure:20_locusts}
}
\subfigure[N = 35]{
\includegraphics[width=0.29\columnwidth]{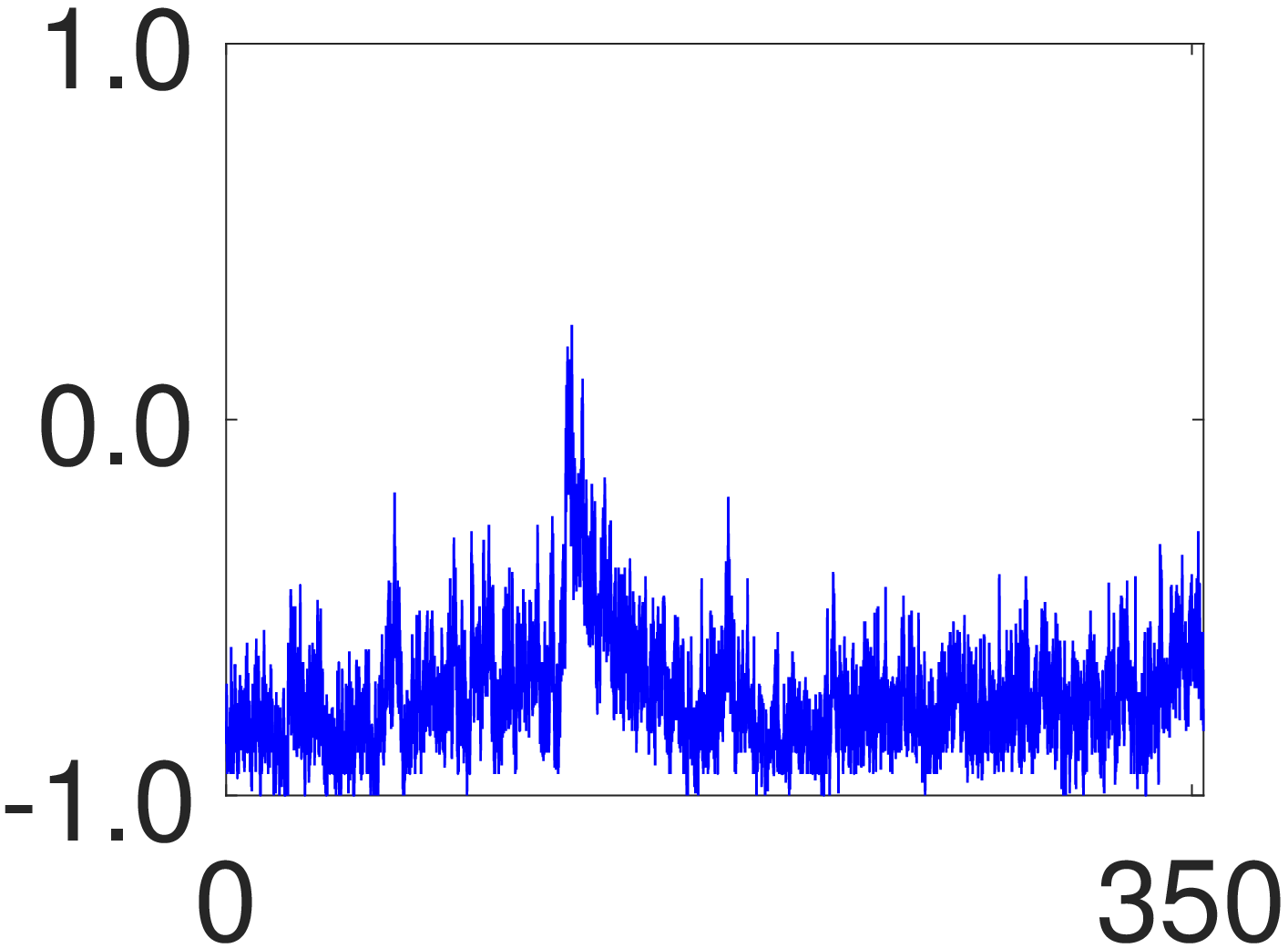}
\label{figure:35_locusts}
}
\subfigure[N = 5]{
\includegraphics[width=0.29\columnwidth]{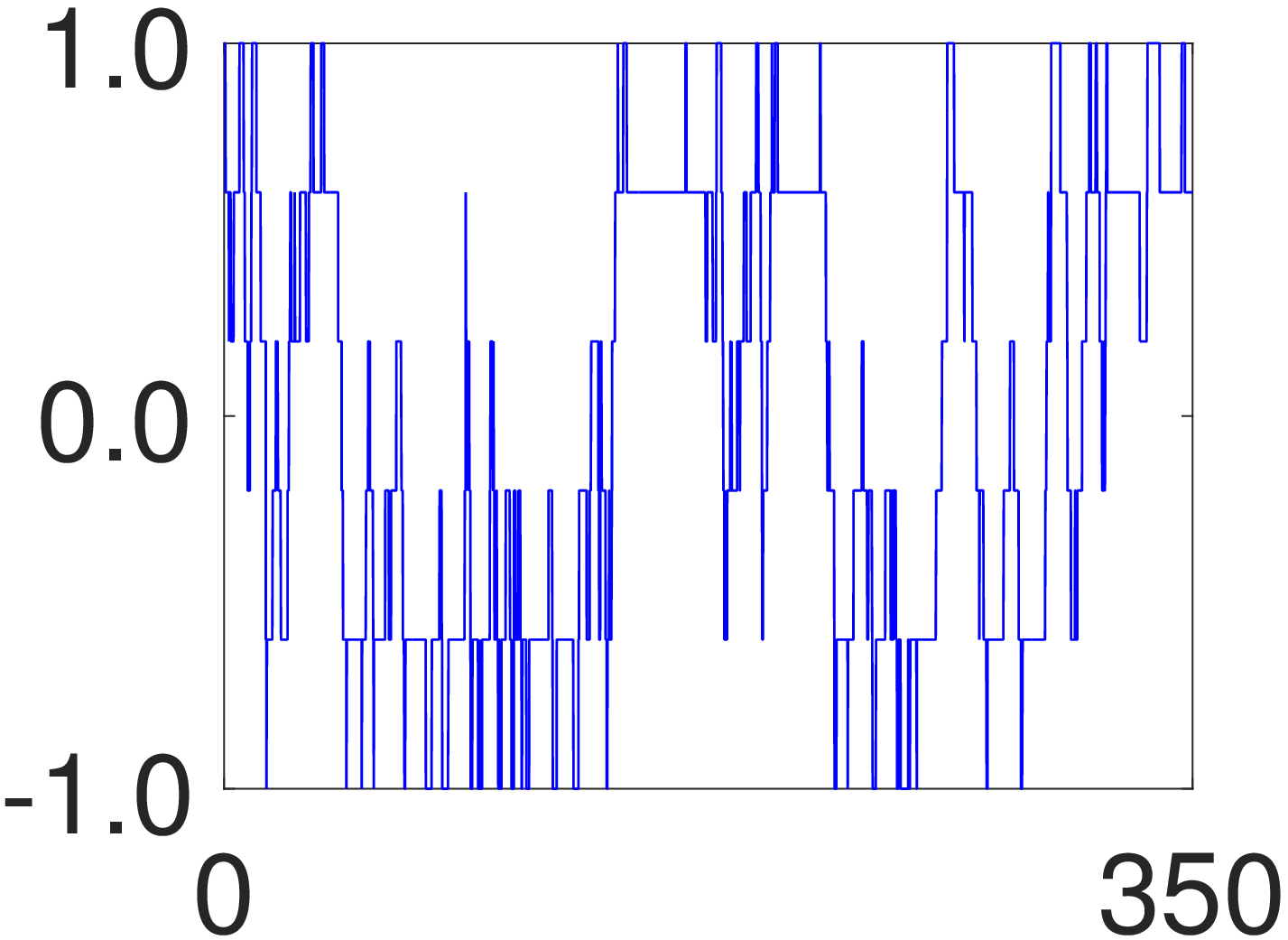}
\label{figure:5_locusts:IBM}
}
\subfigure[N= 20]{
\includegraphics[width=0.29\columnwidth]{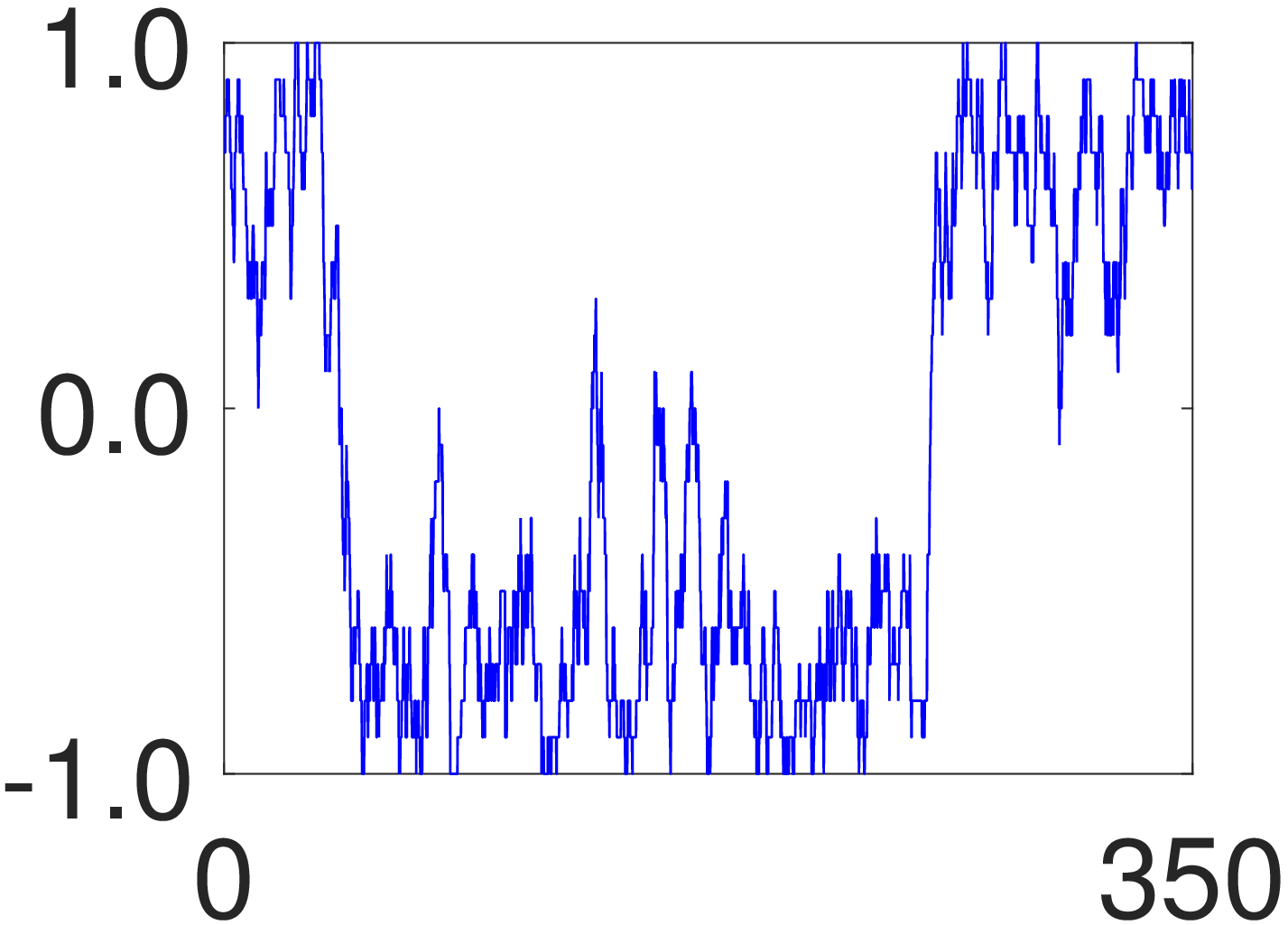}
\label{figure:20_locusts:IBM}
}
\subfigure[N = 35]{
\includegraphics[width=0.29\columnwidth]{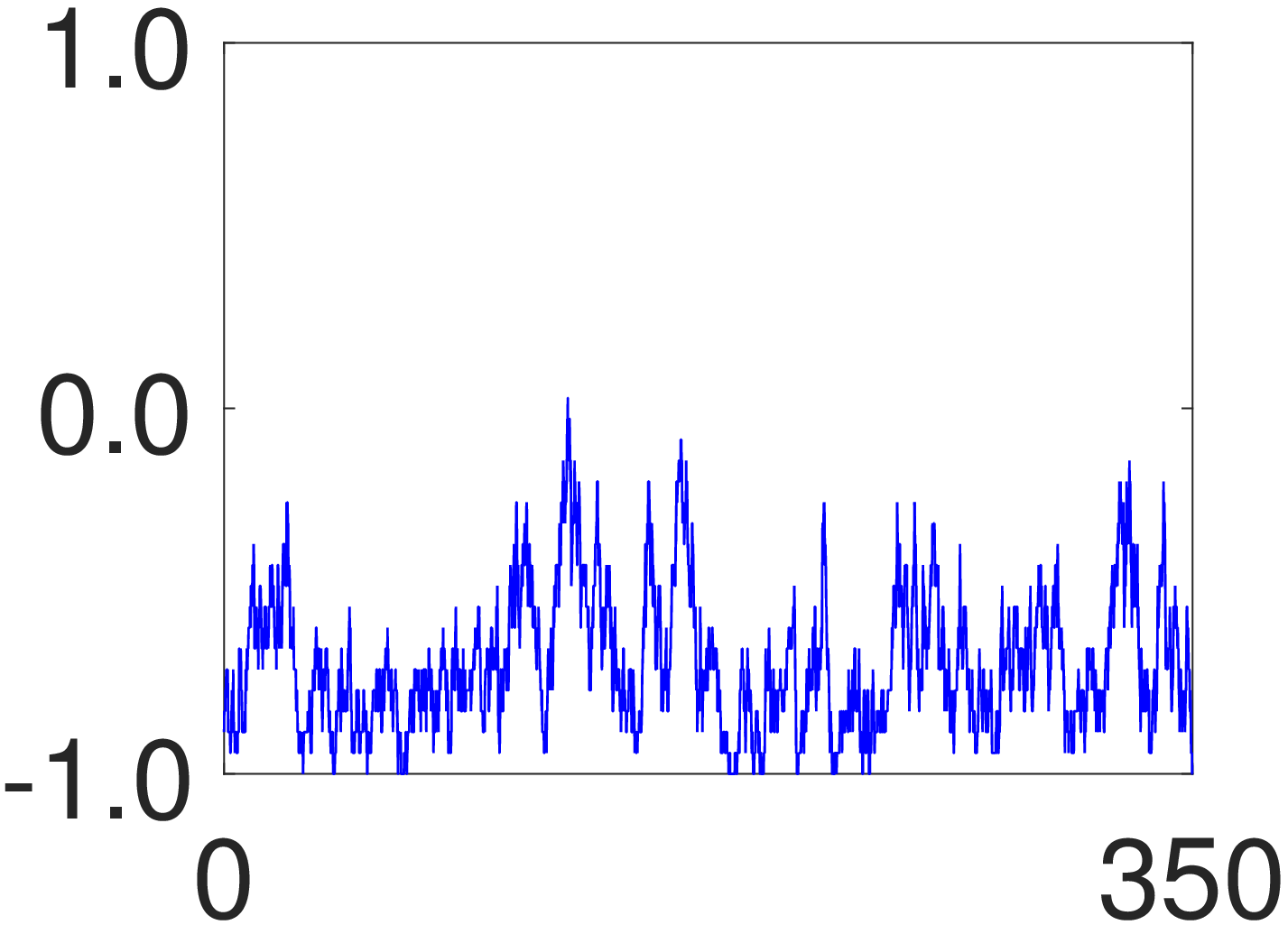}
\label{figure:35_locusts:IBM}
}
\subfigure[N = 5]{
\includegraphics[width=0.29\columnwidth]{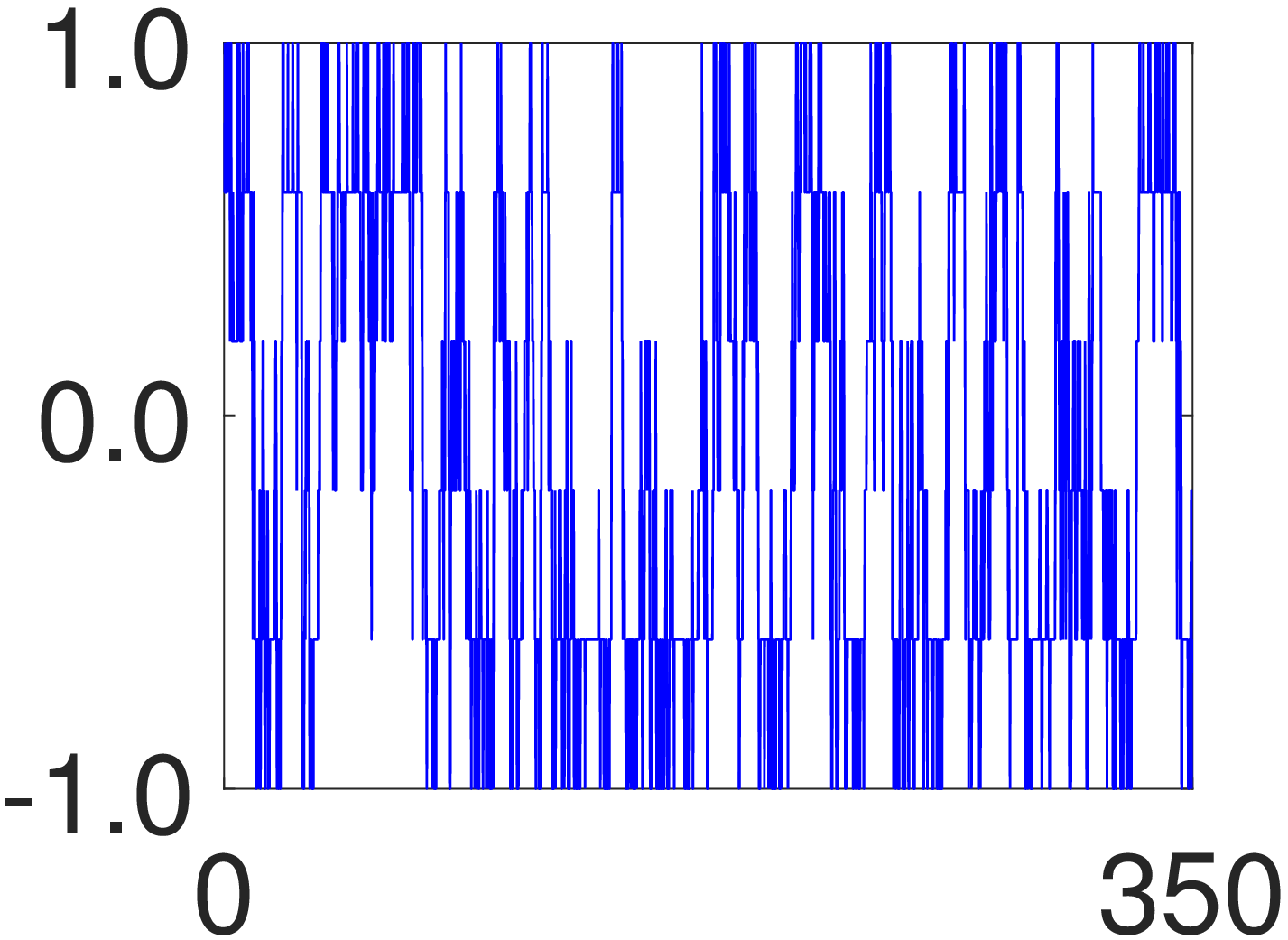}
\label{figure:5_locusts:IBM:av}
}
\subfigure[N= 20]{
\includegraphics[width=0.29\columnwidth]{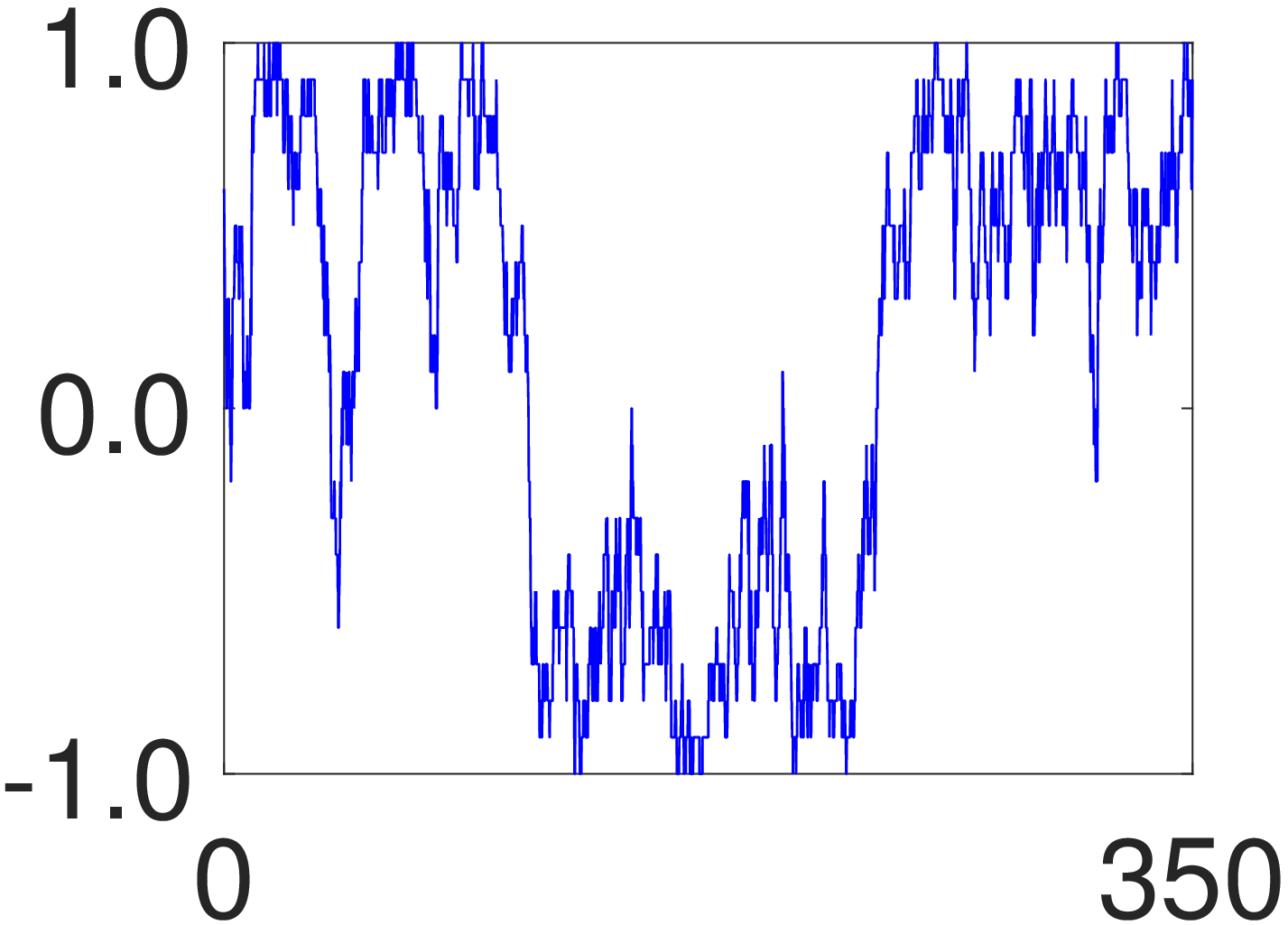}
\label{figure:20_locusts:IBM:av}
}
\subfigure[N= 35]{
\includegraphics[width=0.29\columnwidth]{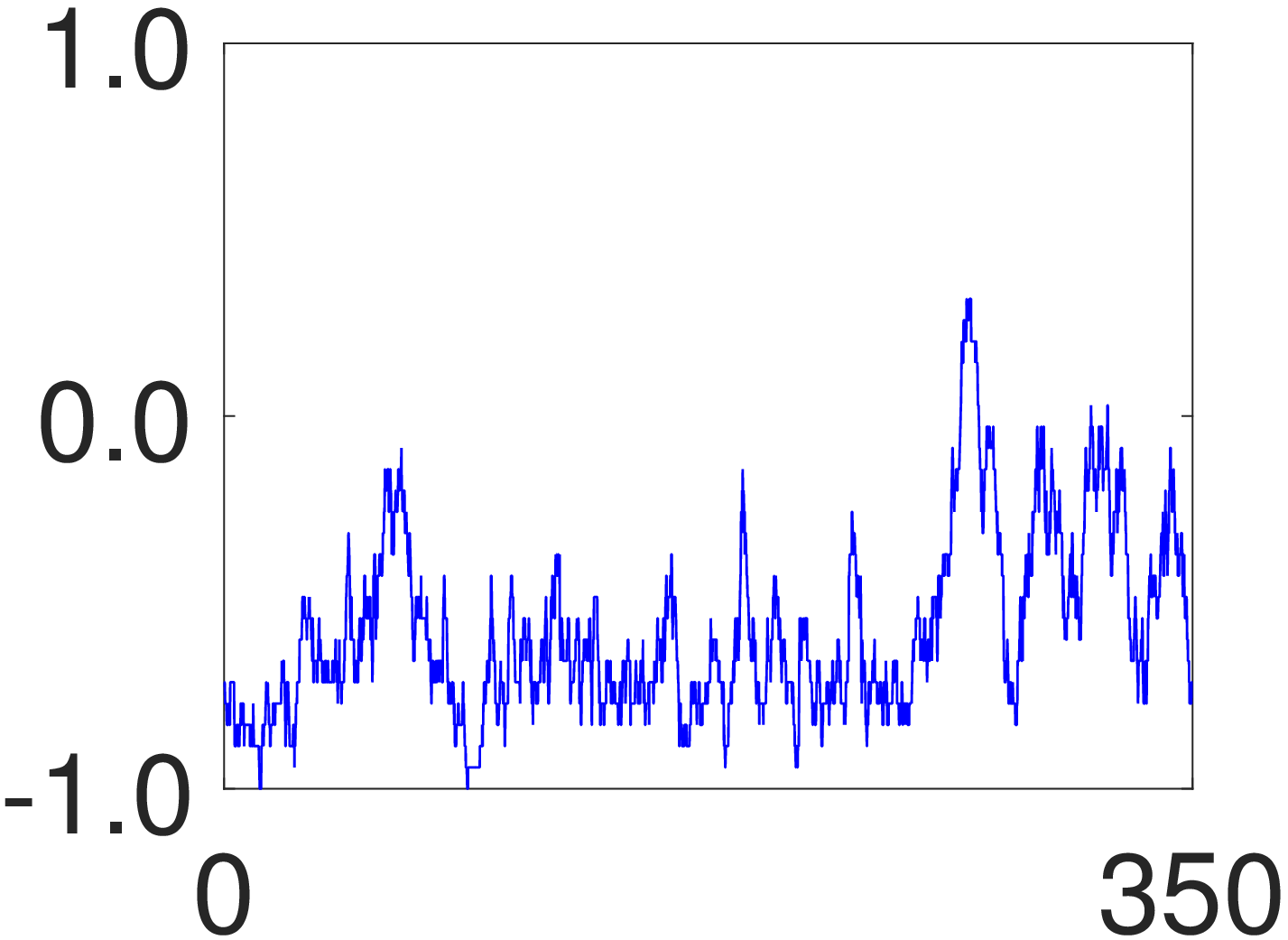}
\label{figure:35_locusts:IBM:av}
}
\end{center}
\caption{(Color online) Change in alignment, $z(t)$, over time demonstrates disorder for 5 locusts (\subref{figure:5_locusts}, \subref{figure:5_locusts:IBM} and \subref{figure:5_locusts:IBM:av}), collective motion with switching between states for 20 locusts (\subref{figure:20_locusts}, \subref{figure:20_locusts:IBM} and \subref{figure:20_locusts:IBM:av}) and more persistent collective motion for 35 locusts (\subref{figure:35_locusts}, \subref{figure:35_locusts:IBM} and \subref{figure:35_locusts:IBM:av}). Experimental data is displayed in \subref{figure:5_locusts}--\subref{figure:35_locusts} and equivalent simulations are shown in \subref{figure:5_locusts:IBM}--\subref{figure:35_locusts:IBM} (using reaction rates, given in Table~\ref{table:Reaction_Rates}, fitted for each value of $N$) and \subref{figure:5_locusts:IBM:av}--\subref{figure:35_locusts:IBM:av} (using reaction rates, $r_1 = 0.0225$, $r_2 = 0.0453$, $r_3 = 0.1664$, fitted to the averaged data (see below)).} 
\label{figure:experimental_data}
\end{figure}

To investigate the effects of population density on the swarming of locusts, Buhl \emph{et al.}~\cite{buhl2006fod} performed a series of experiments, placing different numbers of locusts in a ring-shaped arena. They recorded the alignment $z(t)$, which essentially gives the proportion of anticlockwise-moving individuals ($x_2$) subtracted from the proportion of clockwise-moving individuals ($x_1$). Thus $z = 1$ (or $z=-1$) would indicate that all individuals are moving clockwise (or anti-clockwise, respectively), while $z = 0$ represents equal numbers of clockwise-moving and anticlockwise-moving individuals. The authors observed a rapid transition from disordered to ordered movement as the group size was increased. At low population densities movement is highly disordered (Fig. \ref{figure:experimental_data}\subref{figure:5_locusts}). At intermediate densities the population displays long periods of coherent marching in one direction, punctuated by occasional fast changes in direction (Fig. \ref{figure:experimental_data}\subref{figure:20_locusts}). At high densities no direction changes can be seen during the experiment (Fig. \ref{figure:experimental_data}\subref{figure:35_locusts}).

Using an equation-free method~\cite{kevrekidis2003efc,erban2006grn,yates2009inc} Yates \emph{et al.}~\cite{yates2009inc} numerically derived the drift and diffusion coefficients of an assumed underlying stochastic differential equation (SDE) from the experimental data of Buhl \emph{et al.}~\cite{buhl2006fod}. The diffusion coefficient was found to be smaller when the locusts were more aligned (i.e. close to $z =\pm 1$). The authors adapted a self-propelled particle (SPP) model~\cite{czirok1999cms} to include this effect and demonstrated that the adapted model displayed qualitatively similar population-level behaviour to the experimental data. Subsequently Bode \emph{et al.}~\cite{bode2010mne} proposed another SPP model incorporating particle attraction as well as alignment. This model inherently generated qualitatively similar drift and diffusion coefficients.

Recently Biancalani \emph{et al.}~\cite{biancalani2014nib} used an individual-based model (IBM) to describe bistability in foraging ant colonies. This model demonstrates a kind of bistability where the intrinsic system noise does not simply cause transitions between stable states present in the deterministic formulation, but instead actively constructs the states themselves. In particular, using a model with two types of individual, who may recruit individuals of the opposing type, or change type at random, the intrinsic noise present in the system is found to be greatest when there are equal numbers of each type of individual and at a minimum when one or other type of individual dominates the population. The authors analytically derive an SDE from the IBM, in which the diffusion coefficient is reduced at the extremes of the domain, similarly to the diffusion coefficient found by Yates \emph{et al.}~\cite{yates2009inc} from the data of Buhl \emph{et al.}~\cite{buhl2006fod}.

In this paper we formulate a minimal model that describes the locust experiment~\cite{buhl2006fod}, following the approach employed by Biancalani \emph{et al.}~\cite{biancalani2014nib}. Using a variant of the Kramers-Moyal expansion \cite{gardiner2009hsm}, we analytically derive an SDE directly from this model and thus give formulas for the explicit dependence of the drift and diffusion coefficients on the total number of individuals. These coefficients indicate that in order to match the experimental data, model locusts must effectively interact with at least two neighbours simultaneously. Using a revised coefficient estimation approach we can also derive drift and diffusion coefficients for the experimental data and we use these to estimate model reaction rates consistent with the experimental data. Interestingly, we find that it is not necessary to explicitly incorporate space in our model in order to reproduce the experimentally derived coefficients. This suggests that the switching behaviour of the locusts is an inherent property of the way they interact with each other and the frequency of those interactions, rather than being a consequence of the particular spatial geometry of the arena. Thus the effect is driven by the density of locusts and their individual interactions. We quantitatively compare our model against the experimental data by deriving the stationary probability distributions (SPDs) and mean first passage times (MFPTs) between clockwise and anticlockwise coherent movement.

\section{Coarse graining the individual-based model}
We consider a population of $N$ individuals, split into clockwise-moving ($X_1$) and anticlockwise-moving ($X_2$) populations. Individuals may change direction spontaneously, or decide to change direction as the result of interactions with one or two locusts travelling in the opposite direction. The model may be summarised in the following system of interactions:
\begin{eqnarray}
X_1\stackrel{\hat{r}_1}{\rightarrow} X_2&,\quad& X_2 \stackrel{\hat{r}_1}{\rightarrow} X_1,\label{first_order}\\
X_1+X_2 \stackrel{\hat{r}_2}{\rightarrow} 2X_1&,\quad& X_1+X_2 \stackrel{\hat{r}_2}{\rightarrow} 2X_2,\label{second_order}\\
2X_1+X_2 \stackrel{\hat{r}_3}{\rightarrow} 3X_1&,\quad& X_1+2X_2 \stackrel{\hat{r}_3}{\rightarrow} 3X_2.\label{third_order}
\end{eqnarray}
Thus the rate of transitioning from state $b$ to state $a$, $T(a\vert b)$ is given by
\begin{eqnarray}
 T^+(x_1)&\equiv& T\left(x_1+\frac{1}{N}\Big|x_1\right) = \sum_{i=1}^3 r_i x_1^{i-1}(1-x_1), \label{equation:transition_increase}\\
 T^-(x_1)&\equiv& T\left(x_1-\frac{1}{N}\Big|x_1\right) = \sum_{i=1}^3 r_i x_1(1-x_1)^{i-1}, \label{equation:transition_decrease}
\end{eqnarray}
where we have rescaled the rates $r_i=\hat{r}_i/N^{i}$ for $i=1,2,3$ when converting between locust numbers, $X_1$, and locust proportions, $x_1 = X_1/N$. Using these transition rates we can write down the master equation for the probability density function $P(x_1,t)$~\cite{van2007spp}:
\begin{equation}\label{me}
	\D{P}{t}(x_1, t) = \sum_{x'_1\ne x_1} \left[ T(x_1| x'_1) P(x'_1,t) - T(x'_1 | x_1)  P(x_1,t) \right].
\end{equation}
Introducing the step operators, $\varepsilon^{\pm}$, which represent the creation or destruction of an individual of species $X_1$ we can Taylor expand in $1/N$, the inverse of the population size \cite{van2007spp}:
\begin{equation}\label{expansion}
	\varepsilon^{\pm} f(x_1) = f(x_1 \pm \frac{1}{N}) \approx \left(1 \pm \frac{1}{N}\partial_{x_1} + \frac{1}{2 N^2}\partial_{x_1}^2\right) f(x_1),
\end{equation}
where $f(x_1)$ is a general function of the fraction of the species, $x_1$. The master equation~(\ref{me}) can be rewritten using the step operators and subsequently approximated using Eq.~(\ref{expansion}) to give
\begin{equation}
	\begin{split}
		\D{P}{t}(x_1,t) =& \left[\left(\varepsilon^- -1 \right) T^+ + \left(\varepsilon_1^+ -1 \right)T^- \right] P(x_1,t) \\
		\approx &  -\frac{1}{N} \D{}{x_1}\left[\left(T^+ - T^-\right)P(x_1,t)\right] \\
		&+ \frac{1}{2 N^2} \DD{}{x_1}\left[\left(T^+ + T^-\right) P(x_1,t)\right],
	\end{split}
\end{equation}
neglecting terms of $\mathcal{O}(1/N^3)$.

Rescaling time using $t/N \to t$ and inserting the expressions for the transition rates (Eqs. (\ref{equation:transition_increase}) and (\ref{equation:transition_decrease})) gives the Fokker-Planck equation
\begin{equation}
	\D{P}{t}(x_1,t) =  - \D{}{x_1} \left[ \mathcal A P(x_1,t)\right] + \frac{1}{2 N} \DD{}{x_1}\left[ \mathcal B P(x_1,t)\right]\label{equation:general_FPE_for_z},
\end{equation}
where $\mathcal A = r_1(1-2x_1)+r_3 x_1 (1-x_1) (2x_1 -1)$ and $\mathcal B = r_1 + 2r_2 x_1 (1-x_1) + r_3 x_1 (1-x_1)$. Or, in terms of $z=2x_1-1$,
\begin{equation}
	\D{P}{t}(z,t) =  - \D{}{z} \left[F(z) P(z,t)\right] + \DD{}{z} \left[D(z) P(z,t)\right],
\end{equation}
for 
\begin{eqnarray}
F(z) &=& -2 r_1 z + r_3 z(1-z^2) / 2,\label{equation:model_drift}\\
D(z) &=& 2(r_1 + (2r_2 + r_3)(1-z^2)/4)/N \label{equation:model_diffusion}.
\end{eqnarray}

This FPE corresponds to the It\^{o} SDE
\begin{equation}
\dot{z} = F(z) + \sqrt{2D(z)}\eta(t)\label{equation:SDE_for_z}, 
\end{equation}
where $\eta(t)$ is Gaussian white noise with zero mean and correlator $\langle\eta(t)\eta(t') \rangle=\delta(t-t')$.

For $r_3 = 0$ this gives the model studied by Biancalani \emph{et al.}\cite{biancalani2014nib}, which displays bistability at for small populations but not for large populations. In contrast, in our system as $N$ increases in size, so that $F(z)$ is the dominant term in the equation, there is an additional pair of non-zero steady states at $z=\pm\sqrt{1-4r_1/r_3}$ in the analogous deterministic system. Hence higher-order interactions between locusts (\emph{i.e.} the $r_3$ interaction) are required in order for the model to demonstrate the observed coherent motion: long periods of clockwise or anticlockwise movement. Note that including higher-order interactions does not change the qualitative population-level phenomena observed here.

\section{Equation-free coefficients}
To quantitatively compare the model with data we estimate the value of the coefficients ($F(z)$ and $D(z)$) using a modified version of the equation-free method used by Yates \emph{et al.}~\cite{yates2009inc} (see supplementary material\footnote{See Supplementary Material at [URL will be inserted by publisher] for more details on deriving the drift and diffusion coefficients from simulations and data.} for a more detailed description of the implementation of the equation-free technique in this context). The modified method makes use of the symmetry of the system, since we do not expect fundamental differences between clockwise-moving and anticlockwise-moving locusts, and requires initial preprocessing (as used in Refs. \citep{buhl2006fod,yates2009inc}) to smooth the data. The initial preprocessing used is a moving time-average with a window of two seconds, and is required to avoid the method becoming overwhelmed with high frequency oscillations, that likely arise from problems in video tracking of the individual locusts. 

\begin{figure}[h!!!!!!!!!!!!!]
\begin{center} 
\subfigure[N = 5]{
\includegraphics[width=0.45\columnwidth]{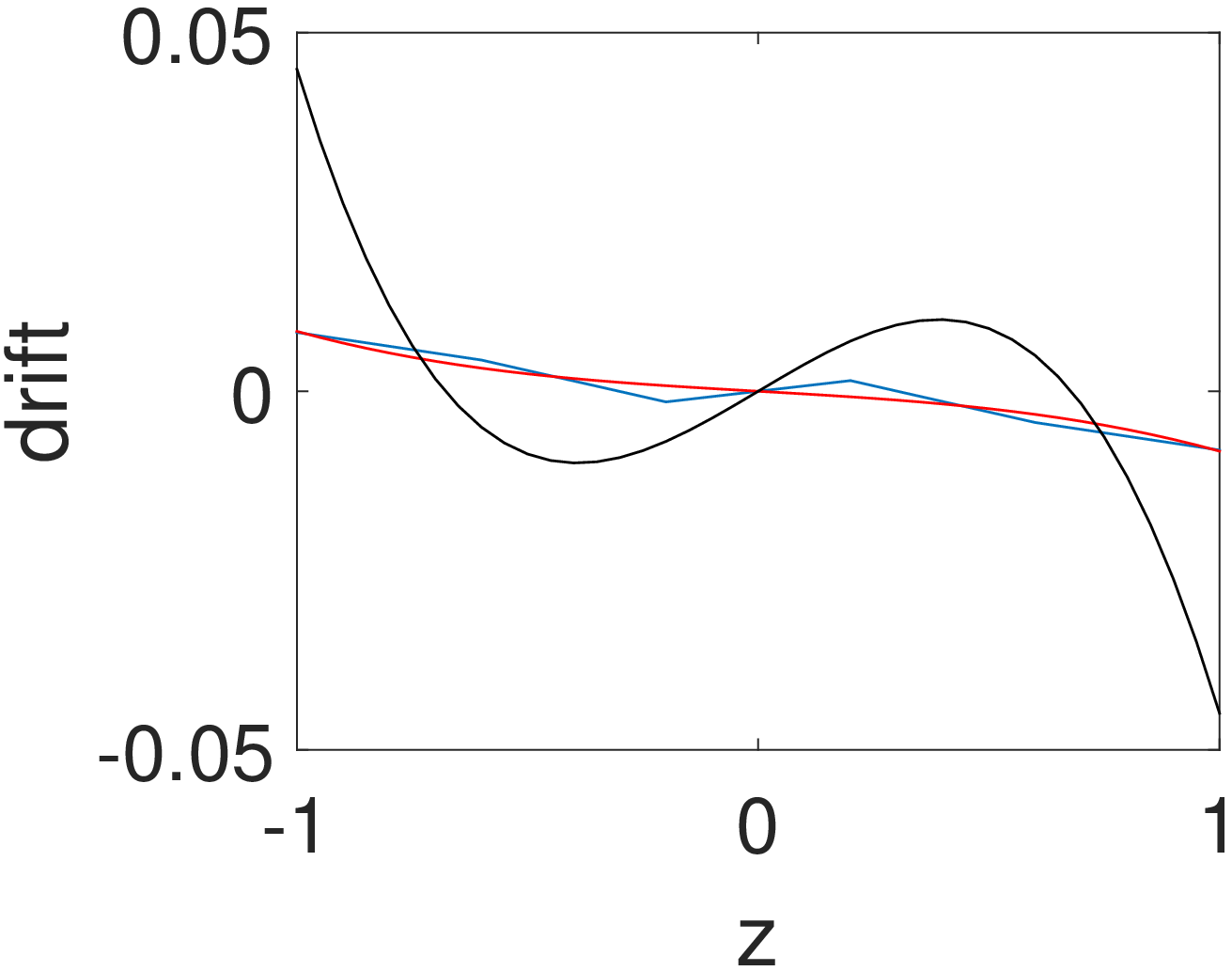}
\label{figure:drift_5_locusts}
}
\subfigure[N = 5]{
\includegraphics[width=0.45\columnwidth]{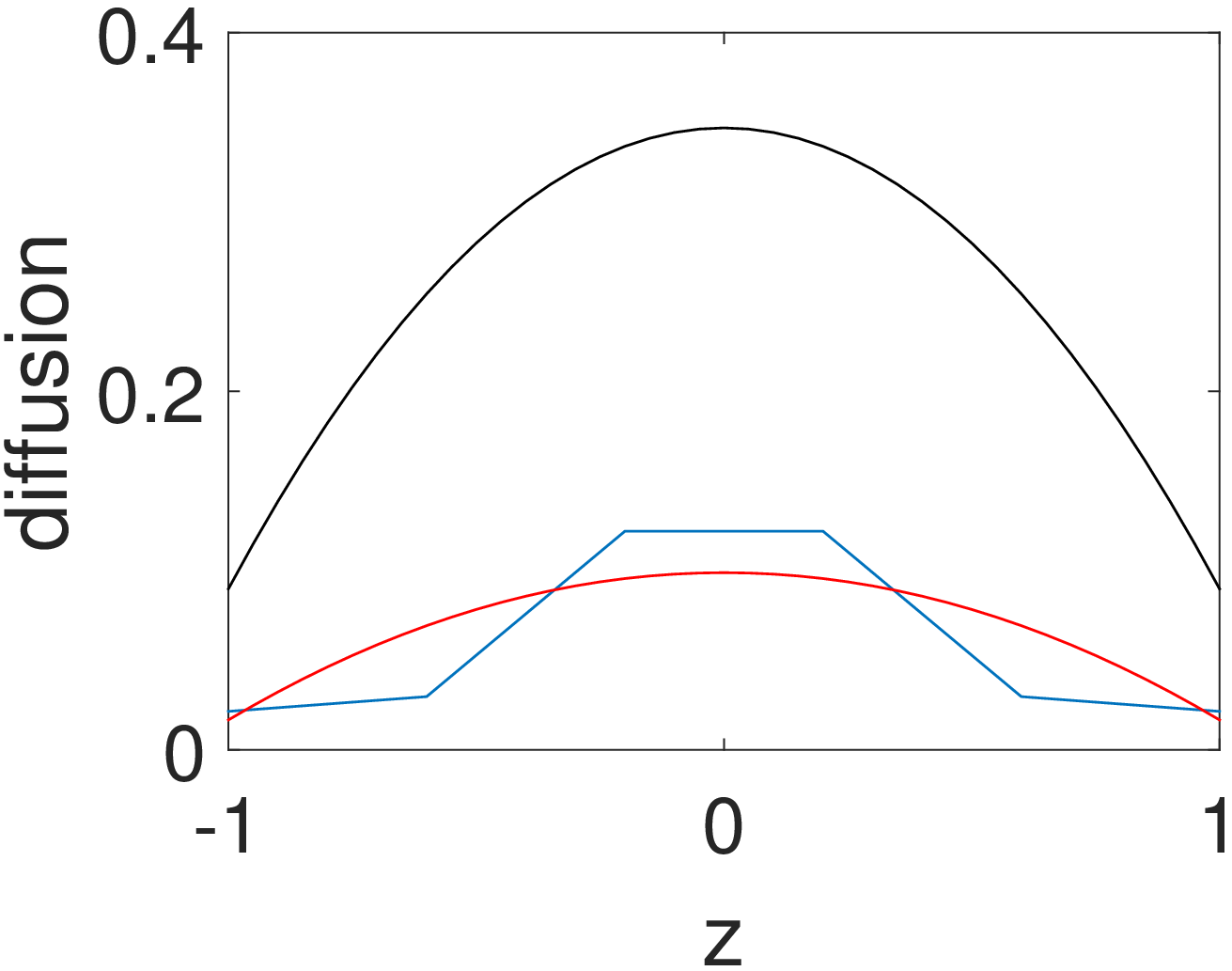}
\label{figure:diff_5_locusts}
}
\subfigure[N= 20]{
\includegraphics[width=0.45\columnwidth]{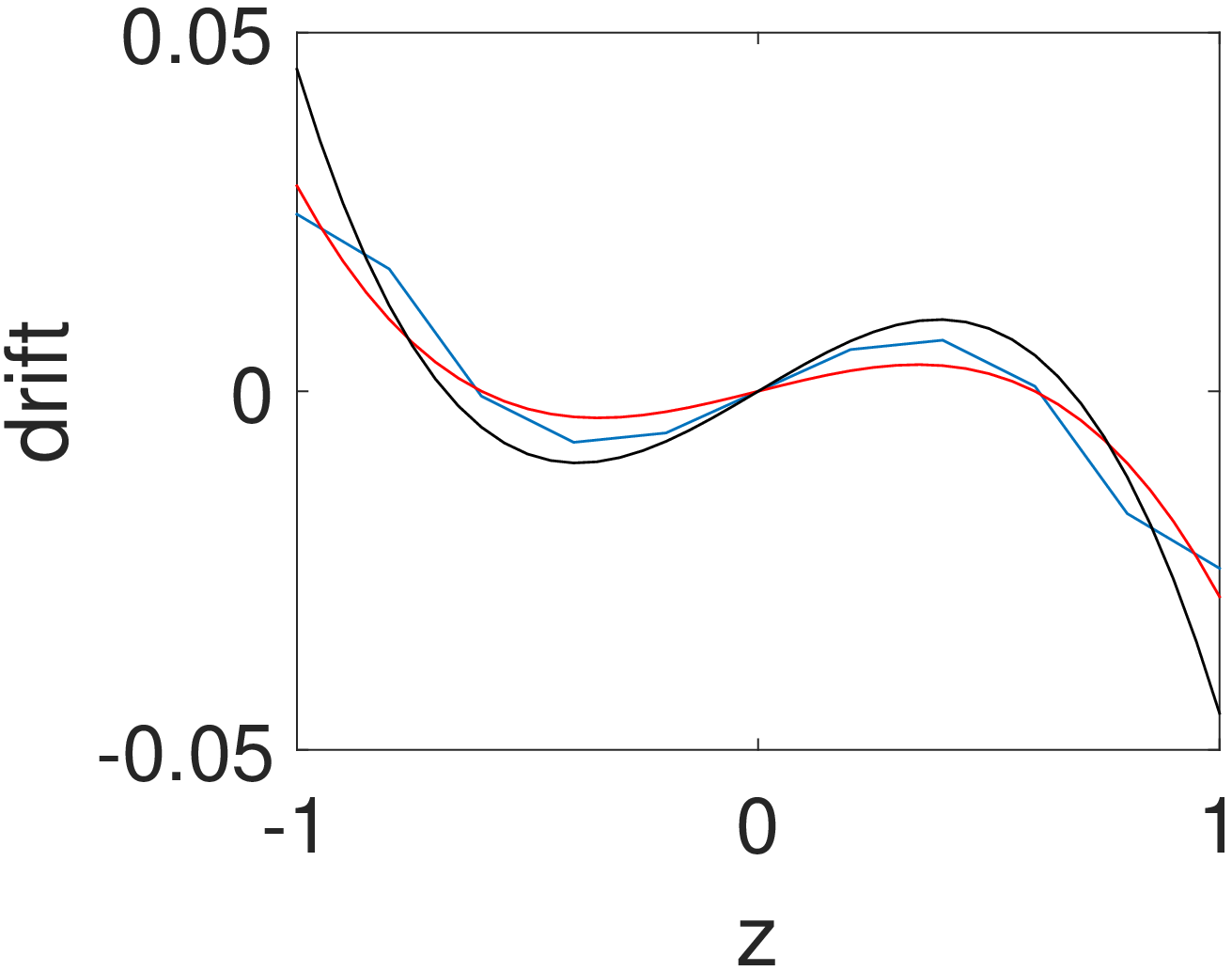}
\label{figure:drift_20_locusts}
}
\subfigure[N = 20]{
\includegraphics[width=0.45\columnwidth]{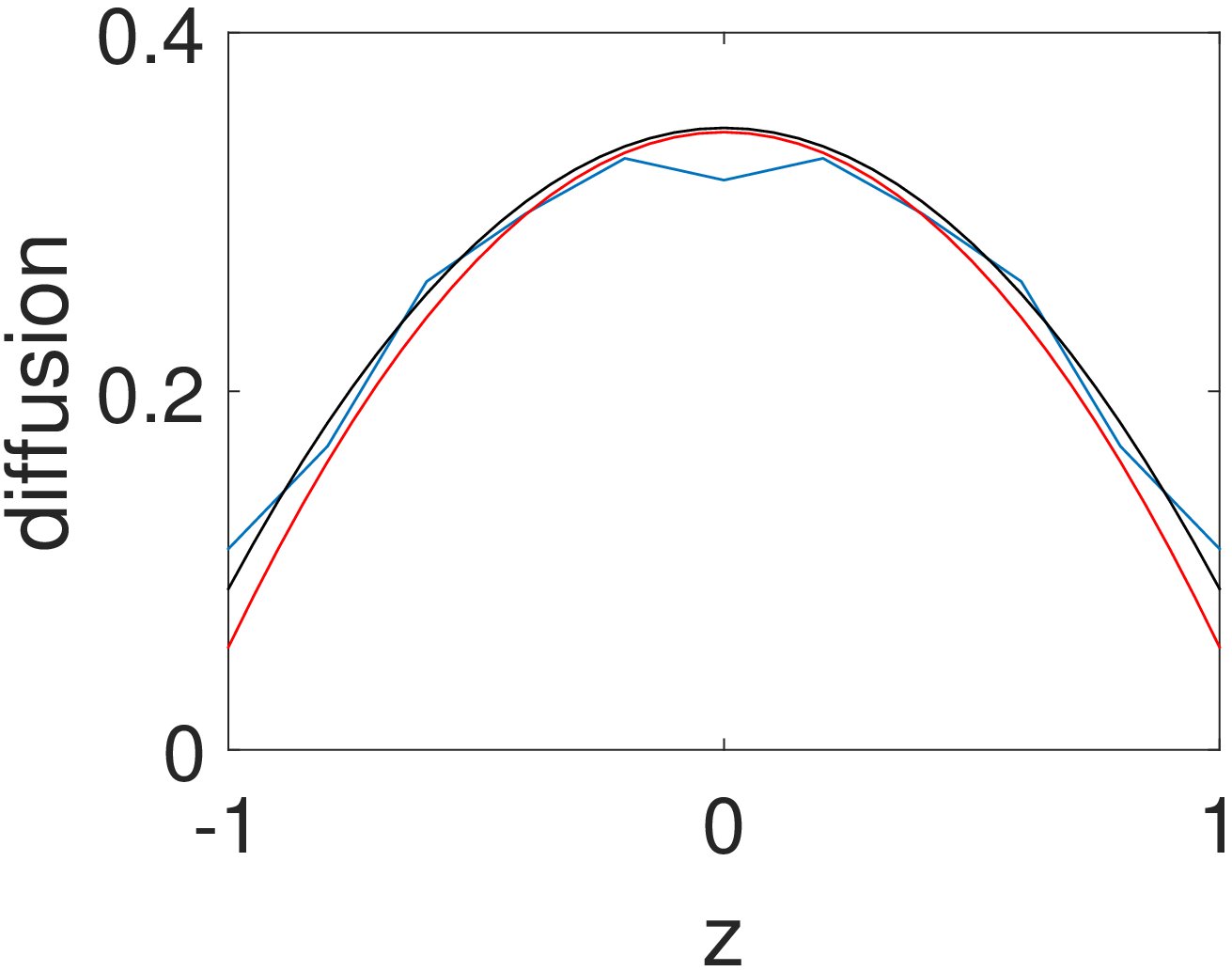}
\label{figure:diff_20_locusts}
}
\subfigure[N = 35]{
\includegraphics[width=0.45\columnwidth]{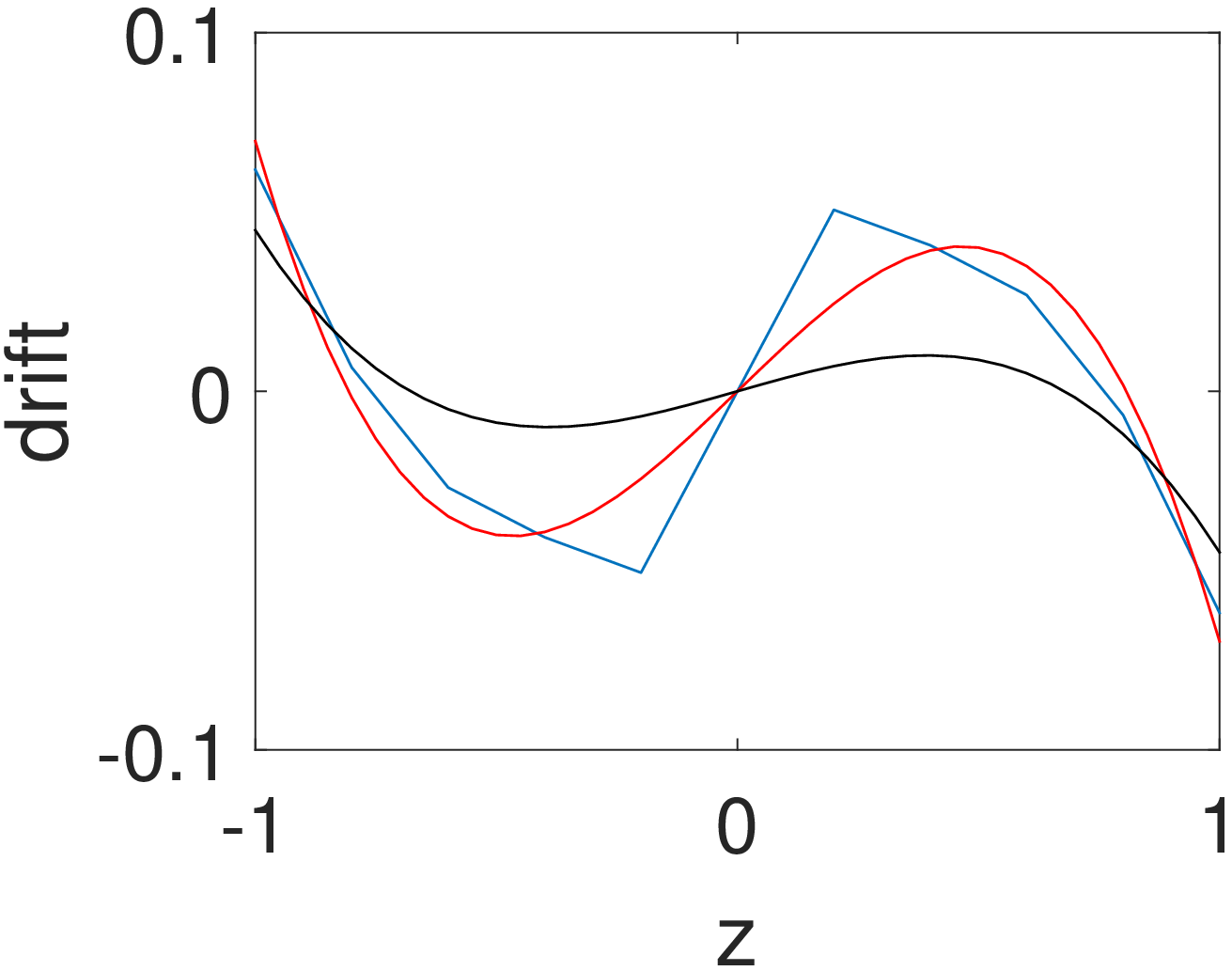}
\label{figure:drift_35_locusts}
}
\subfigure[N = 35]{
\includegraphics[width=0.45\columnwidth]{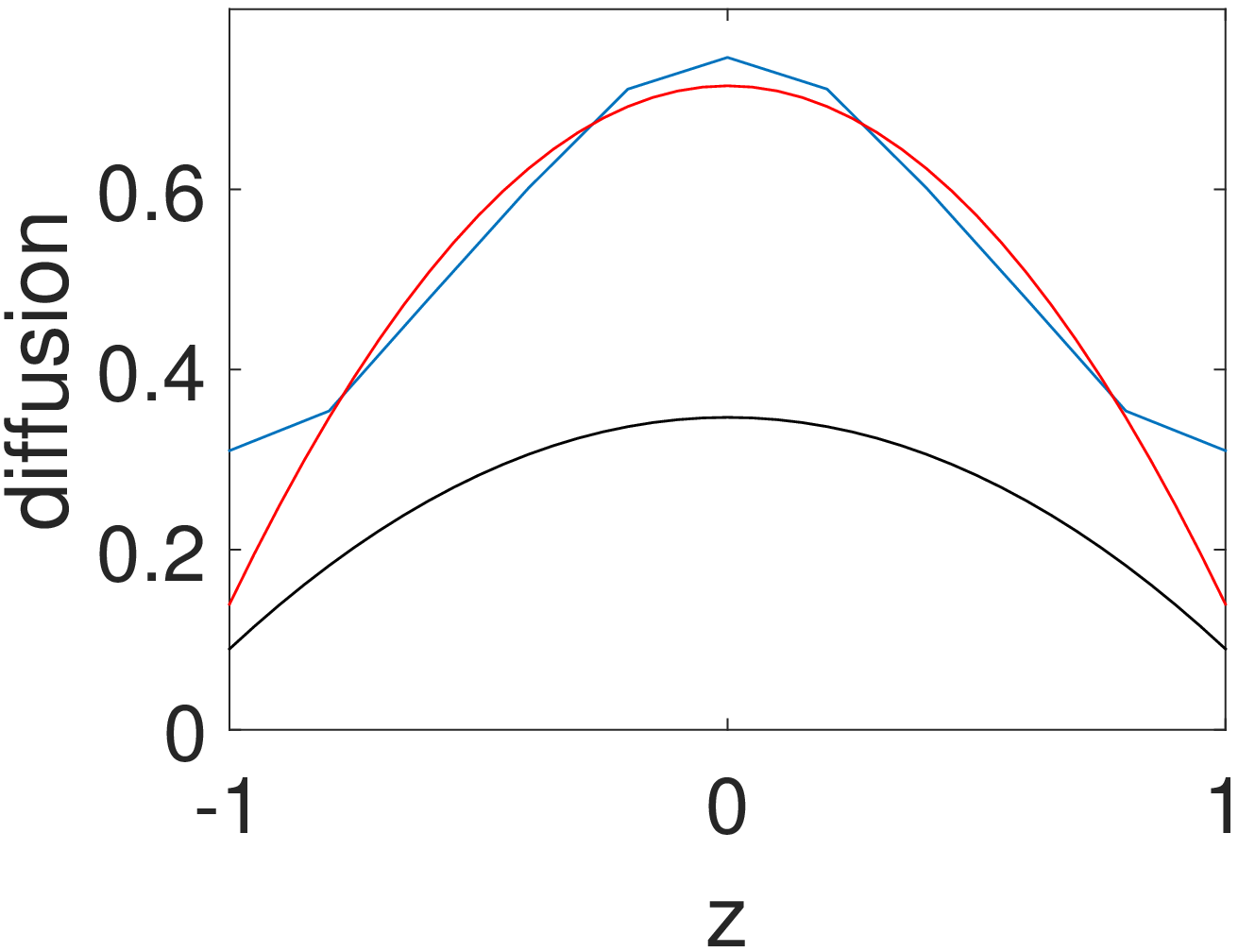}
\label{figure:diff_35_locusts}
}
\end{center}
\caption{(Color online) Experimentally derived drift ($F(z)$, \subref{figure:drift_5_locusts}, \subref{figure:drift_20_locusts} and \subref{figure:drift_35_locusts}) and diffusion coefficients ($D(z)$, \subref{figure:diff_5_locusts}, \subref{figure:diff_20_locusts} and \subref{figure:diff_35_locusts})compared to those generated by our model for 5 (\subref{figure:drift_5_locusts} and \subref{figure:diff_5_locusts}), 20 (\subref{figure:drift_20_locusts} and \subref{figure:diff_20_locusts}) and 35 locusts (\subref{figure:drift_35_locusts} and \subref{figure:diff_35_locusts}). In each figure the noisy blue (light gray) curve is the experimentally derived coefficient, the red (smooth light gray) curve uses individually fitted parameters values (see Table~\ref{table:Reaction_Rates}) and the black curve uses fitted parameter values averaged over all the experiments (see Fig.~\ref{figure:experimental_data} for rates.)}
\label{figure:equation_free}
\end{figure}

For each group size, we find that the estimated diffusion (Fig.~\ref{figure:equation_free}\subref{figure:diff_5_locusts}--\subref{figure:diff_35_locusts}) and drift coefficients (Fig.~\ref{figure:equation_free}\subref{figure:drift_5_locusts}--\subref{figure:drift_35_locusts}) are consistent with the functional forms found by our analysis. We estimate the parameter values, $r_1$ to $r_3$, by fitting Eqs.~(\ref{equation:model_drift}) and (\ref{equation:model_diffusion}) to the equation-free derived coefficients using the non-negative least squares method (see Appendix \ref{section:Fitting_the_estimated_coefficients}).
The value of the interaction rates resultant from the least squares fitting are given in Table \ref{table:Reaction_Rates}. 

\begin{table}
\begin{tabular}{l|rrrrrrrrrr}
$N$ & 5    & 6   &  7   & 10   & 15   & 20  &  25 &   30   & 35    & 40 \\
\hline
  $r_1$ &  0.004  &  0.009 &   0.006  &  0.009  &  0.025  &  0.014  &  0.016   & 0.031  &  0.035   & 0.042  \\
$r_2$ &    0.036   & 0.011  &  0.046  &  0.073  &       0.000   & 0.099  &  0.189&    0.011   & 0.088  &  0.102 \\
$r_3$ &    0.009  &  0.017 &   0.002   & 0.022  & 0.143   & 0.090  &  0.064  &  0.289   & 0.400  &  0.413 
\end{tabular}
\caption{Parameter values fitted for each value of $N$ (see supplementary material).}
\label{table:Reaction_Rates}
\end{table}

Allowing the parameter values to vary with $N$ allows an extremely good fit to the data (Fig. \ref{figure:equation_free}, red lines) and simulating the IBM with these parameter values gives a good qualitative agreement to the original time series data (Fig.~\ref{figure:experimental_data}\subref{figure:5_locusts:IBM}--\subref{figure:35_locusts:IBM}). We may also rescale the experimental data by $N$, so that we can fit all experiments together to give one `average' value for each of the parameters $r_1$ to $r_3$. This gives a less good fit to the data (Fig.~\ref{figure:equation_free}, black lines), but still shows reasonable agreement with far fewer total parameters. The average parameters also give a good qualitative agreement to the time series data (Fig.~\ref{figure:experimental_data}\subref{figure:5_locusts:IBM:av}--\subref{figure:35_locusts:IBM:av}). 
Comparisons for a wider range of values of $N$ are given in supplementary material Fig.~S2.

In both cases the equation-free-generated parameter values capture the large-scale switching behaviour, showing disorder at small populations sizes (Fig.~\ref{figure:experimental_data}\subref{figure:5_locusts}, \subref{figure:5_locusts:IBM} and \subref{figure:5_locusts:IBM:av}), long periods of coherent motion, switching between $z=\pm 1$ at intermediate population sizes (Fig.~\ref{figure:experimental_data}\subref{figure:20_locusts}, \subref{figure:20_locusts:IBM} and \subref{figure:20_locusts:IBM:av}), and sustained clockwise or anticlockwise movement of the population at high densities (Fig.~\ref{figure:experimental_data}\subref{figure:35_locusts}, \subref{figure:35_locusts:IBM} and \subref{figure:35_locusts:IBM:av}). The high-frequency fluctuations are not captured by this technique, due to the necessity of smoothing the initial data and the sensitivity of the equation-free method to the degree of `discreteness' in the underlying data. This sensitivity arises from the underlying assumption when deriving the FPE (and thus when deriving the equation-free method) that $z$ is a continuous variable. This assumption is clearly more valid at higher population densities, however we have shown that the large-scale population dynamics are still well-captured even at lower densities. 

\section{Stationary probability distributions}
To test the quantitative fit of our model more systematically, we calculate the SPD.
This is found analytically by setting $\partial P(z,t)/\partial t=0$ in Eq.~(\ref{equation:general_FPE_for_z}) and solving the resulting ordinary differential equation (ODE):
\begin{equation}
\DDstraight{}{z}\left[D(z)P_s(z)\right]-\Dstraight{}{z}\left[F(z)P_s(z)\right]=0.\label{equation:simplified_SDE_for_z}
\end{equation} 
This can simply be integrated once with respect to $z$ to leave us with the first order ODE
\begin{equation}
\Dstraight{}{z}\left[D(z)P_s(z)\right]-F(z)P_s(z)=C.\label{equation:simplified_SDE_for_z_integrated}
\end{equation} 
The constant of integration $C$ is set to zero (assuming there are no sources or sinks of probability) leaving us with a homogeneous first order ODE which can be solved by means of an integrating factor: $\exp(-\int \{F(z)/D(z)\}\ud z)$ to give
\begin{equation}
\begin{split}
P_s(z)&=\\
&c\left[4r_1+(2r_2+r_3)(1-z^2)\right]^{\frac{4Nr_1(r_2+r_3)}{(2r_2+r_3)^2}-1} e^{\frac{r_3z^2N}{2(2r_2+r_3)}},
\end{split}
\label{equation:analytical_SPD}
\end{equation}
where $c$ is a normalisation constant for the probability density function.

\begin{figure}[h!!!!!]
\begin{center} 
\includegraphics[width=0.5\columnwidth]{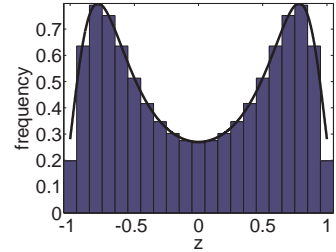}
\end{center}
\caption{(Color online) The stationary probability distribution derived analytically and calculated by long-run individual-based simulations for $N=20$ locusts. The black curve corresponds to the analytical solution (see equation \eqref{equation:analytical_SPD}) and the blue (light gray) histograms to the results of the simulation. In both scenarios we take reaction rates, $r_1 = 0.0225$, $r_2 = 0.0453$, $r_3 = 0.1664$, which correspond to the rates derived from fitting our analytical expressions for drift and diffusion to the averaged data.}
\label{figure:comparison_of_SPDs}
\end{figure}

In order to corroborate our theoretically derived stationary probability distribution we have carried out an individual-level simulation. By recording the alignment values at appropriately spaced time points we have determined a simulation-based SPD  with which we compare our analytically derived SPD in Fig. \ref{figure:comparison_of_SPDs}. 

\begin{figure}[h!!!!!!!!]
\begin{center} 
\subfigure[N = 5]{
\includegraphics[width=0.46\columnwidth]{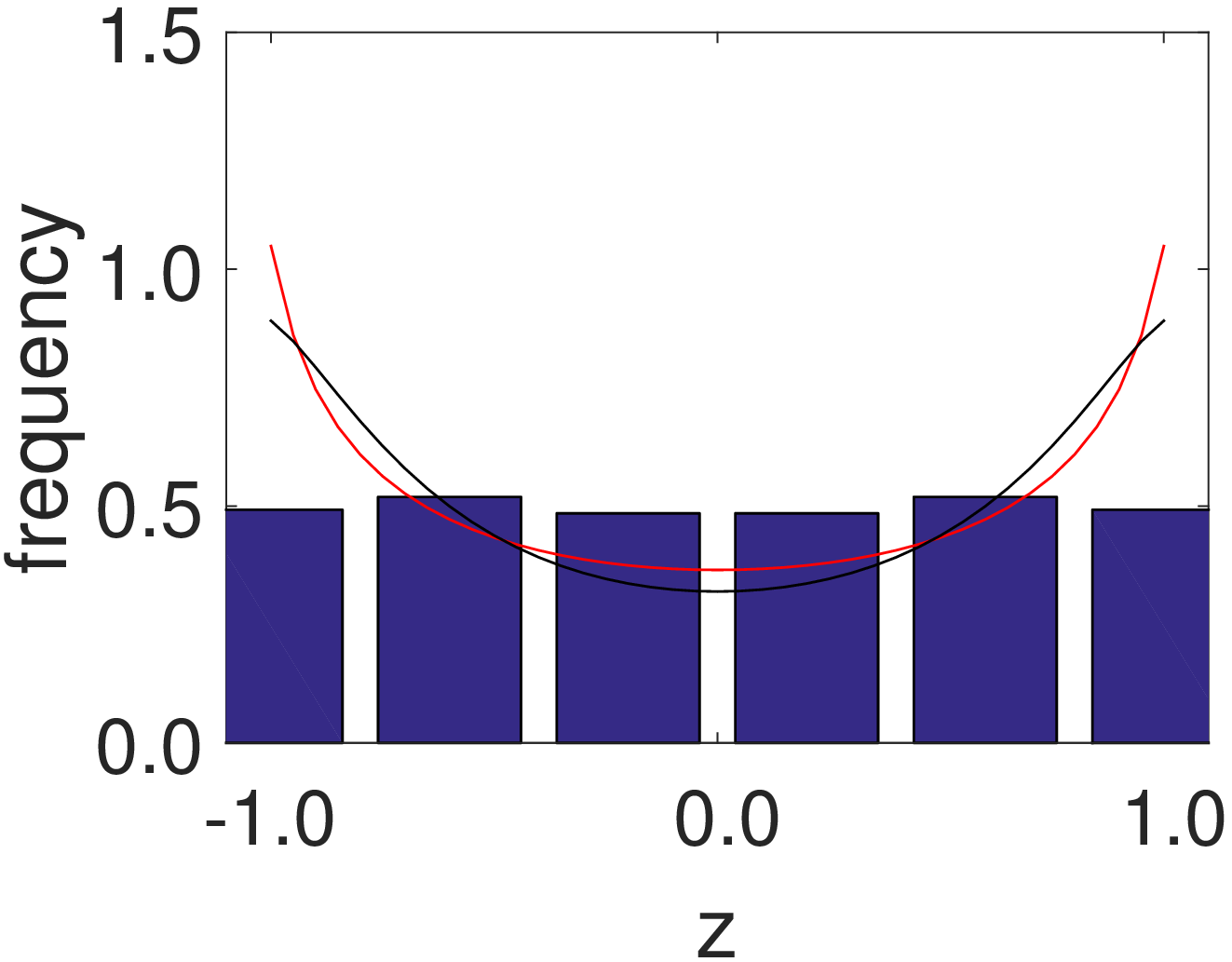}
\label{figure:SPD:5_locusts}
}
\subfigure[N= 20]{
\includegraphics[width=0.46\columnwidth]{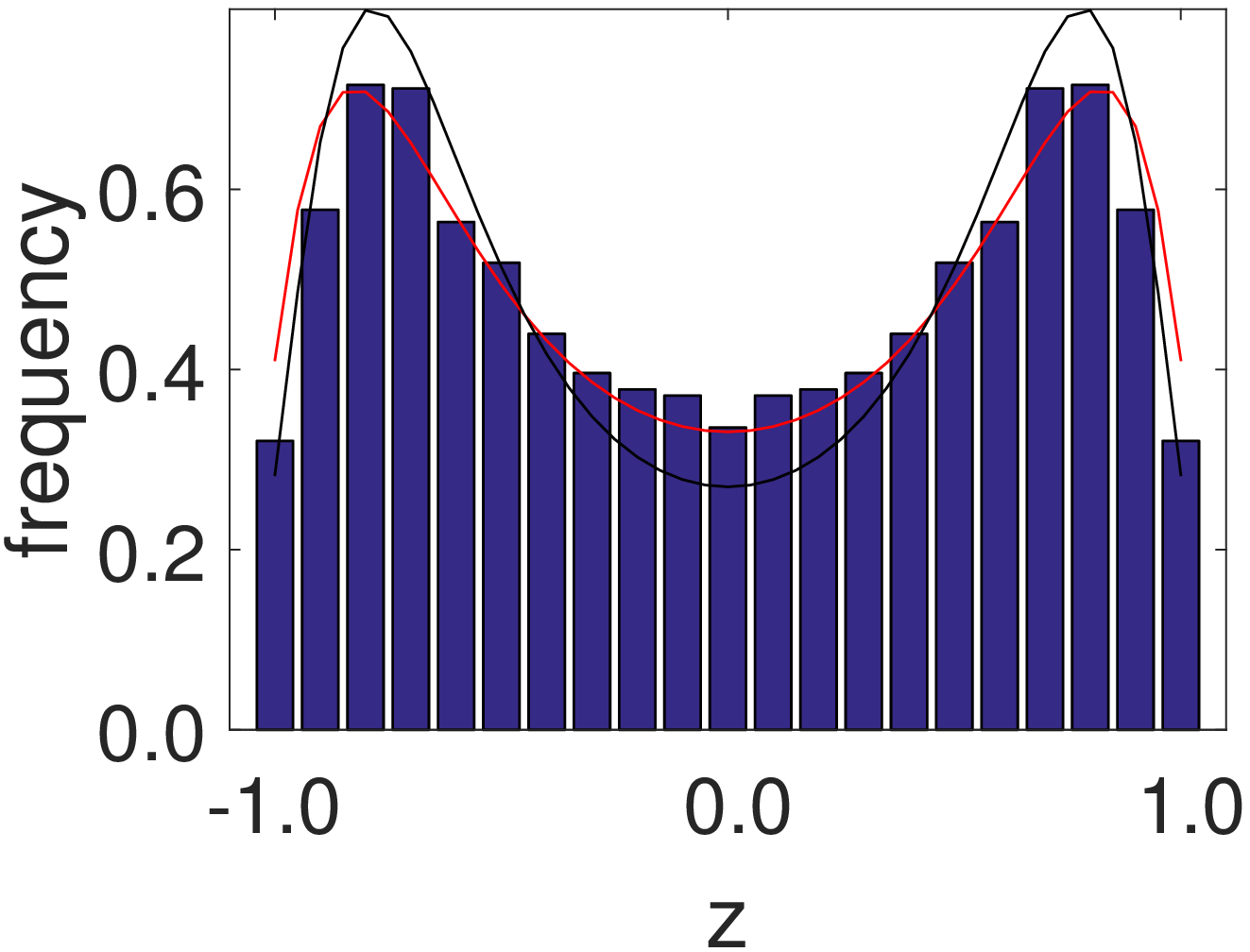}
\label{figure:SPD:20_locusts}
}
\subfigure[N = 35]{
\includegraphics[width=0.46\columnwidth]{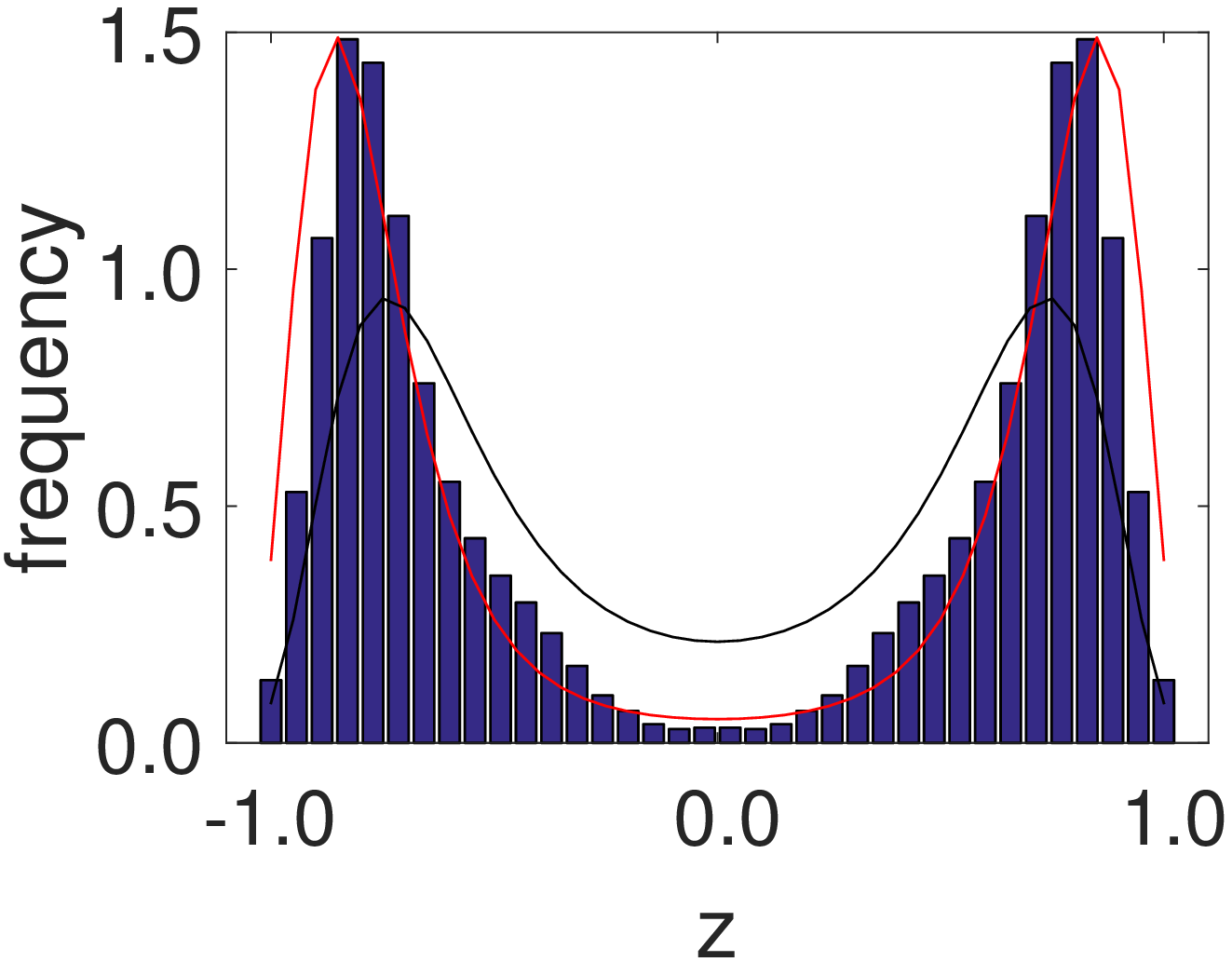}
\label{figure:SPD:35_locusts}
}
\end{center}
\caption{(Color online) Experimentally observed stationary probability distributions (bars) compared to analytically predicted distributions with fitted parameters (red (light gray) lines) and averaged fitted parameters (black lines) for \subref{figure:SPD:5_locusts} 5 , \subref{figure:SPD:20_locusts} 20 and \subref{figure:SPD:35_locusts} 35 locusts. See Table \ref{table:Reaction_Rates} and Fig.~\ref{figure:experimental_data} for reaction rates.}
\label{figure:SPD}
\end{figure}

In a similar manner we may also compare this analytically derived SPD with the fitted parameters to the histograms of the experimental data for each value of $N$ (Fig.~\ref{figure:SPD}). As before, we show the analytical result for both the reaction rates fitted for each $N$ (red lines) and for the `averaged' reaction rates (black lines). The analytical result fits well in both cases, with the least good fit at the lowest population size (Fig.~\ref{figure:SPD}\subref{figure:SPD:5_locusts}), where the assumption that $z$ is a continuous variable is least reasonable. Fig.~\ref{figure:SPD} shows the progression from an disordered population, to one that spends the majority of the time with most individuals moving in the same direction. Comparisons for a wider range of values of $N$ are given in supplementary material Fig.~S3.

\section{Mean first passage times}
\begin{figure}[h!!!!!!!!!!]
\begin{center} 
\subfigure[]{
\includegraphics[width=0.46\columnwidth]{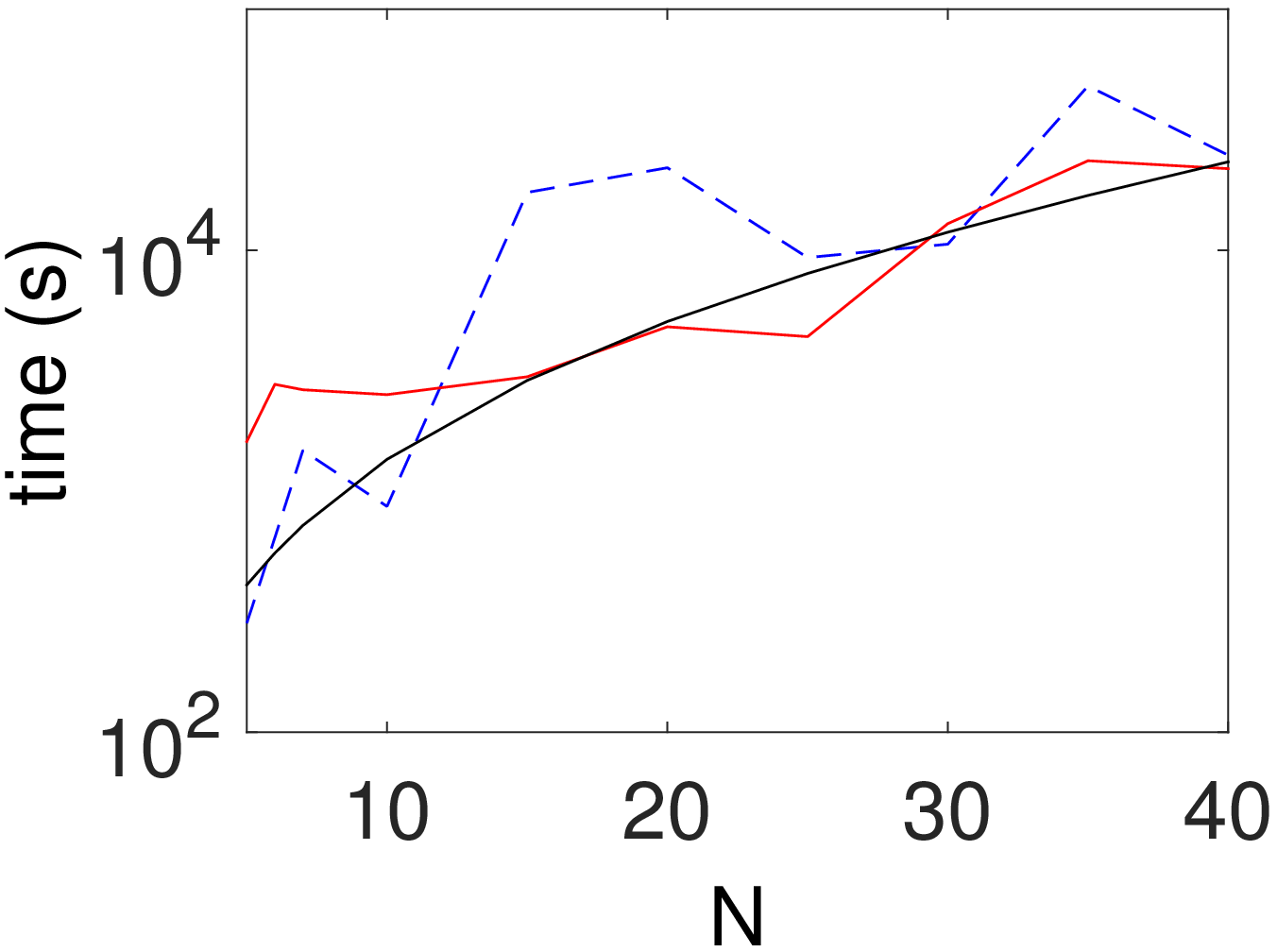}
\label{figure:MFPT}
}
\subfigure[]{
\includegraphics[width=0.46\columnwidth]{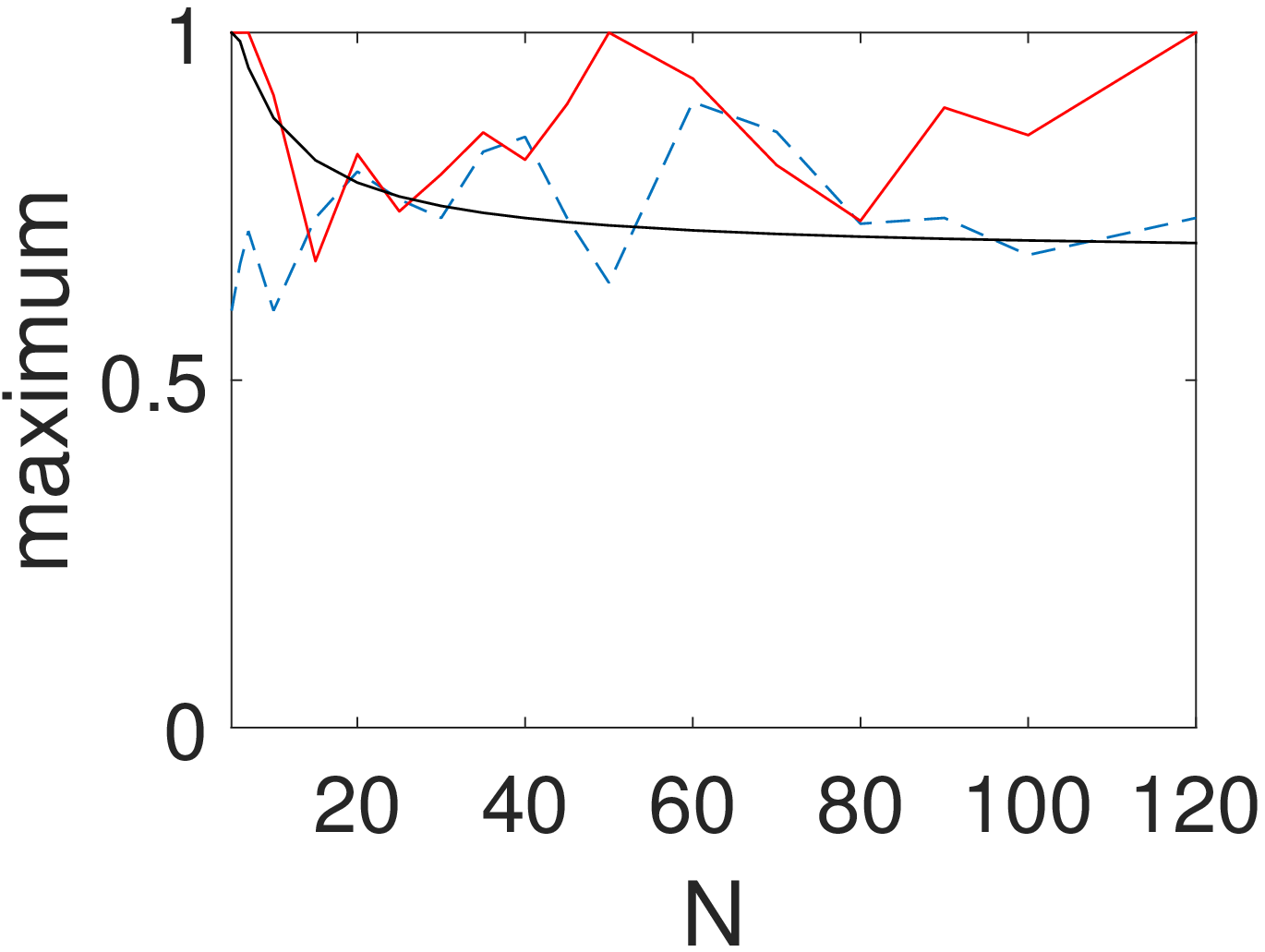}
\label{figure:max_pos}
}
\end{center}
\caption{(Color online) \subref{figure:MFPT} Experimentally observed mean first passage time (dotted blue line) compared to analytically predicted distributions with individually fitted parameters (red line) and averaged fitted parameters (black line) for a range of values of $N$. \subref{figure:max_pos} Position of maximum ($z_m$) for experimental data (dotted blue line), and the model with individually fitted parameters (red line) and averaged fitted parameters (black line) for a range of values of $N$. See Table \ref{table:Reaction_Rates} and Fig.~\ref{figure:experimental_data} for reaction rates.}
\label{figure:MFPT_and_max_pos}
\end{figure}

The time courses generated by our IBM seem to display similar periods of time spent in each state as those found in the experimental data (Fig.~\ref{figure:experimental_data}). To compare these quantitatively we calculate the MFPT between the two maximum values of the SPD, found numerically for the data. The maxima are given analytically (using Eq.~\eqref{equation:analytical_SPD}) as
\begin{equation}
z = \pm z_m = \pm\sqrt{\frac{4r_2 + 2r_3 + N(r_3-4r_1)}{Nr_3}}.
\end{equation}
To find the MFPT~\cite{gardiner2009hsm} we then solve
\begin{equation}
F(z)\Dstraight{T}{z}+D(z)\DDstraight{T}{z}=-1 \label{equation:mean_first_passage_time},
\end{equation}
numerically, subject to a reflecting boundary condition ($\ud T / \ud z = 0$) at $z = -1$ and an absorbing boundary at the position of the positive maximum ($T(z_m)=0$). Equation~(\ref{equation:mean_first_passage_time}) is derived from the backwards FPE, which gives the occupancy probability conditioned on the initial position (see Appendix \ref{section:derivation_of_the_mean_first_passage_time}).
The average time taken for the system to move from $z=-z_m$ to $z=z_m$ is then given by $T(-z_m)$. The MFPT and the position of the maxima are shown in Fig.~\ref{figure:MFPT_and_max_pos}. Analytical predictions fit the general trend of the experimental data well, demonstrating that the demographic noise present in our model successfully reproduces the density-dependent effects seen in the data. We note that the experimental SPDs are very flat at low population densities (as seen in Fig.~\ref{figure:SPD:5_locusts}), which may account for the discrepancy in maximum positions seen in Fig.~\ref{figure:max_pos}.

\section{Discussion}
In this paper we have introduced a minimal model to describe the onset of cohesive motion of a group of locusts as the group size increases. We demonstrate that collective behaviour can be initiated through simple individual-based interactions, and show that there is an explicit dependence on the size of the group considered. Our model implies that it is necessary to include third-order interactions between locusts (\emph{i.e.} non-zero $r_3$), in order to have the directional coherence at large population sizes that is generated by the exponential term in Eq.~(\ref{equation:analytical_SPD}).
We note that this model will display switches between clockwise-moving and anti-clockwise moving populations for a wide range of parameter values and is generic in systems of this kind.

We have tested our model quantitatively against experimental data, by first using the equation-free method \cite{yates2009inc} to fit for the parameter values $r_1$ to $r_3$, and then comparing our analytic predictions against the SPD and MFPT found experimentally as the number of individuals varies. We note that, as has been studied for the model by Biancalani \emph{et al.}~\cite{houchmandzadeh2015efn,saito2014tad}, it is possible to consider the master equation directly, to derive exact, but complicated, formulas. These do not add to our intuition about the model. While our model does not replicate exactly the high frequency oscillations found in the original data, it does capture large-scale population-level behaviours such as the existence of coherent steady states near $z=\pm1$ and the timescale of switching between these states. 

\medskip

\appendix


\section{Fitting the estimated coefficients}\label{section:Fitting_the_estimated_coefficients}
Using a least squares formulation it is possible to fit the model parameters $r_1,r_2$, and $r_3$ in order to simultaneously match the drift and diffusion coefficients of the model to those of the data. Denoting the discretised forms of the experimentally-derived drift and diffusion coefficients by the vectors $\bs{F}=(F_0\dots F_M)$ and $\bs{D}=(D_0\dots D_M)$ and the discrete alignment vector $\bs{z}=(z_0,\dots,z_M)$, the appropriate formulation of the least squares problem is as follows: 
\begin{equation}
 \begin{bmatrix}
  F_0 \\ \vdots \\ F_M \\ D_0 \\ \vdots \\ D_M
 \end{bmatrix} = 
\begin{bmatrix}
 -z_0 & 0 & z_0(1-z_0^2)/4 \\
 -z_1 & 0 & z_1(1-z_1^2)/4 \\
\vdots & \vdots & \vdots \\  
 -z_M & 0 & z_M(1-z_M^2)/4 \\
1/2 & (1-z_0^2)/4 & (1-z_0^2)/8 \\
1/2 & (1-z_1^2)/4 & (1-z_1^2)/8 \\
\vdots & \vdots & \vdots \\
1/2 & (1-z_M^2)/4 & (1-z_M^2)/8 \\
\end{bmatrix}
\begin{bmatrix}
r_1 \\ r_2 \\ r_3 
\end{bmatrix}.
\end{equation}

Averaged rates are found by a similar method, but we first scale the diffusion coefficient $D(z)$ by the number of locusts, $N$, so that all the experimental data may be used together to find just one set of averaged rates. Fig.~S2 of the supplementary material demonstrates comparison between the experimentally derived drift and diffusion coefficients and the fitted coefficients produced by our model for a range of values of $N$.
We have the additional constraint that none of our rates can be negative which requires us to employ non-negative least squares \cite{lawson1995}. We solve the least squares problem for each value of $N$ using the active set algorithm as implemented in \texttt{MATLAB}'s non-negative least-squares optimiser \texttt{lsqnonneg}. 

Note that although our model does not explicitly incorporate space, space is implicitly taken account of by our reactions rates.

\section{Derivation of the mean first passage time}\label{section:derivation_of_the_mean_first_passage_time}

We wish to find the mean time taken for a locust swarm completely aligned in one direction to become completely aligned in the opposite direction.
In short we are interested in the mean time for the system, starting at $z=-z_m$ or $z=z_m$ to arrive at $z=z_m$ or $z=-z_m$ respectively. 
Clearly, by employing the individual-based model, we can calculate this quantity by averaging over many appropriately initialised simulations or through one long simulation run, recording the times taken for the system to move from $z\leq -z_m$ to $z\geq z_m$ and \emph{vice versa}. We may also calculate the mean first passage time by employing the coarse-grained version of the model. The method is standard~\cite{gardiner2009hsm}, and we also give it here for completeness.

We begin by considering the backward Fokker-Planck (or Kolmogorov) equation. This describes the evolution of $Q(y,t|z,s)$, the probability of the system having alignment $y$ at time $t$, given that the system was at alignment $z$ at an earlier time, $s$. The backward FPE differs to the forward FPE in that it considers changes with respect to the initial conditions, and is given by 
\begin{equation} \D{Q}{s}(y,t|z,s)=-F(z)\D{Q}{z}(y,t|z,s)-D(z)\DD{Q}{z}(y,t|z,s)\label{equation:backwards_kolmogorov_equation}.
\end{equation}
The probability that the system is still in the region of interest after time $t$, starting at position $z$ is 
\begin{equation} G(z,t) = \int_{-1}^{z_m} Q(y,t|z,0) \ud z,
\end{equation}
and, since the system is time homogeneous, $Q(y,t|z,0) = Q(y,0|z,-t)$, the backward Fokker-Planck equation becomes
\begin{equation}\D{Q}{t}(y,t|z,0)=F(z)\D{Q}{z}(y,t|z,0)+ D(z)\DD{Q}{z}(y,t|z,0).
\end{equation}
Integrating this equation over $y\in[-1,z_m]$, we obtain an evolution equation for the probability that the system remains in the interval $[-1,z_m]$ at time $t$, given the system started at $z\in[-1,z_m]$:
\begin{equation} \D{G}{t}=F(z)\D{G}{z}+  D(z)\DD{G}{z}\label{equation:backwards_kolmogorov_equation_integrated}.
\end{equation}
We must specify the appropriate initial and boundary conditions for this equation. Since we start in the required region at position $z$, we have the initial condition
\begin{equation}
 G(z,0)=1\label{equation:initial_condition}.
\end{equation}
Since, without loss of generality, we are interested in the first exit time at $z=z_m$ we will specify an absorbing boundary condition there
\begin{equation}
 G(z_m,t)=0\label{absorbing_boundary_condition},
\end{equation}
and since there is no flux of probability at $z=-1$ we implement a reflecting boundary~\cite{gardiner2009hsm} there
\begin{equation}
 \D{G}{z}\Big|_{z=-1}=0\label{zero_flux_boundary_condition}.
\end{equation}

Now the probability that the process first leaves $[-1,1]$ is given by $-\partial G / \partial t$ and so the mean time, $T(z)$ for this to happen, as a function of the initial position $z$ is given by
%
%
\begin{eqnarray}
 T(z)&=&-\int^{\infty}_0t\D{G}{t}\ud t,\nonumber\\
 &=&\int^{\infty}_0 G(z,t)\ud t,
\end{eqnarray}
using integration by parts, the initial condition \eqref{equation:initial_condition} and the assumption that all processes will eventually reach $z=1$.

Integrating equation \eqref{equation:backwards_kolmogorov_equation_integrated} (and employing the initial condition \eqref{equation:initial_condition}) and the associated boundary conditions \eqref{absorbing_boundary_condition} and \eqref{zero_flux_boundary_condition} over all time, leaves us with a second-order ordinary differential equation for the mean first passage time
\begin{equation}
F(z)\Dstraight{T}{z}+D(z)\DDstraight{T}{z}=-1 \label{equation:mean_first_passage_time_appendix}
\end{equation}
and boundary conditions
\begin{equation}
\Dstraight{T}{z}\Big|_{z=-1}=0,
\end{equation}
and 
\begin{equation}
 T(z_m)=0,\label{equation:absorbing_boundary_condition_T}
\end{equation}
which specify a well-posed boundary value problem. The mean time for the system to move from $z=-z_m$ to $z=z_m$ is now given by $T(-z_m)$.

\vspace{10pt}
\begin{acknowledgements}
L.D. was supported under EPSRC grant EP/H02171X. J.B. was funded by the Australian Research Council Future Fellowship and Discovery Projects programs.
\end{acknowledgements}

%


\end{document}